\documentclass[11pt,a4paper]{article}
\usepackage{amsmath,amssymb,bm,epsfig,color,graphicx,cite,braket,mathrsfs,slashed,multirow,hyperref}
\newcommand{\lib}[1]{\texttt{#1}}

\renewcommand{\thefootnote}{\fnsymbol{footnote}}

\usepackage{hyperref,xcolor,doi}
\hypersetup{
    colorlinks=true,
    linktocpage=true,
    linkcolor=blue,
    filecolor=blue,      
    urlcolor=blue,
    citecolor=blue,
    }

\begin{document}

\title{
\begin{flushright}
\begin{minipage}{0.2\linewidth}
\normalsize
CTPU-PTC-23-16 \\*[50pt]
\end{minipage}
\end{flushright}
\begin{flushleft}
{\Large \bf 
Constraints on dark matter-neutrino scattering from the Milky-Way satellites and
subhalo modeling for dark acoustic oscillations
\\*[20pt]}
\end{flushleft}
}

\author{
Kensuke Akita$^{1}$\footnote{\href{mailto:kensuke8a1@ibs.re.kr}{kensuke8a1@ibs.re.kr}}
\ \ \ \ and\ \ \ \ Shin'ichiro Ando$^{2,3}$\footnote{\href{mailto:s.ando@uva.nl}{s.ando@uva.nl}}
\\*[20pt]
$^1${\it \small
Particle Theory and Cosmology Group, Center for Theoretical Physics
of the Universe,} \\
{\it \small Institute for Basic Science (IBS),
Daejeon, 34126, Korea} \\
$^2${\it \small
GRAPPA Institute, University of Amsterdam, 1098 XH Amsterdam, The Netherlands} \\
$^3${\it \small
Kavli Institute for the Physics and Mathematics of the Universe (Kavli IPMU, WPI),} \\
{\it \small University of Tokyo, Kashiwa, Chiba 277-8583, Japan} \\*[30pt]
}

\date{
\centerline{\small \bf Abstract}
\begin{minipage}{0.9\linewidth}
\medskip \medskip \small 
The elastic scattering between dark matter (DM) and radiation can potentially explain small-scale observations that the cold dark matter faces as a challenge, as damping density fluctuations via dark acoustic oscillations in the early universe erases small-scale structure.
We study a semi-analytical subhalo model for interacting dark matter with radiation, based on the extended Press-Schechter formalism and subhalos’ tidal evolution prescription.
We also test the elastic scattering between DM and neutrinos using observations of Milky-Way satellites from the Dark Energy Survey and PanSTARRS1.
We conservatively impose strong constraints on the DM-neutrino scattering cross section of $\sigma_{{\rm DM}\text{--}\nu,n}\propto E_\nu^n$ $(n=0,2,4)$ at $95\%$ confidence level (CL), $\sigma_{{\rm DM}\text{--}\nu,0}< 10^{-32}\ {\rm cm^2}\ (m_{\rm DM}/{\rm GeV})$, $\sigma_{{\rm DM}\text{--}\nu,2}< 10^{-43}\ {\rm cm^2}\ (m_{\rm DM}/{\rm GeV})(E_\nu/E_{\nu}^0)^2$ and $\sigma_{{\rm DM}\text{--}\nu,4}< 10^{-54}\ {\rm cm^2}\ (m_{\rm DM}/{\rm GeV})(E_\nu/E_{\nu}^0)^4$, where $E_\nu$ is the neutrino energy and $E_\nu^0$ is the average momentum of relic cosmic neutrinos today, $E_\nu^0 \simeq 6.1\ {\rm K}$.
By imposing a satellite forming condition, we obtain the strongest upper bounds on the DM-neutrino cross section at $95\%$ CL, $\sigma_{{\rm DM}\text{--}\nu,0}< 4\times 10^{-34}\ {\rm cm^2}\ (m_{\rm DM}/{\rm GeV})$, $\sigma_{{\rm DM}\text{--}\nu,2}< 10^{-46}\ {\rm cm^2}\ (m_{\rm DM}/{\rm GeV})(E_\nu/E_{\nu}^0)^2$ and $\sigma_{{\rm DM}\text{--}\nu,4}< 7\times 10^{-59}\ {\rm cm^2}\ (m_{\rm DM}/{\rm GeV})(E_\nu/E_{\nu}^0)^4$.
\end{minipage}
}

\maketitle{}
\thispagestyle{empty}
\addtocounter{page}{-1}
\clearpage
\noindent\hrule
\tableofcontents
\noindent\hrulefill

\renewcommand{\thefootnote}{\arabic{footnote}}
\setcounter{footnote}{0}

\section{Introduction}
\label{sec1}

Dark matter (DM) is the most dominant but elusive matter in the universe. The DM mass and interactions beyond gravity still remain a mystery.
The standard collisionless cold dark matter (CDM) cosmological model is that DM interacts only gravitationally with other particles. While this model explains the cosmic microwave background (CMB) and large-scale structure (LSS) data \cite{Planck:2018vyg,eBOSS:2020yzd}, it has challenges in explaining observations in the small-scale structure \cite{Bullock:2017xww} such as the missing satellite \cite{Klypin:1999uc,Moore:1999nt}, core-cusp \cite{Spergel:1999mh,Walker_2011,Salucci:2011ee,Donato_2009} and too-big-to-fail \cite{Boylan_Kolchin_2011,Papastergis:2014aba} problems. 
Galaxy formation physics may solve these problems \cite{Brooks:2012vi,Read_2017,DES:2015zwj,Arraki_2013}, but observations on small scales still provide a chance to explore DM scenarios beyond the CDM cosmological model. 

Interactions of DM beyond gravity with the Standard Model (SM) (i.e., photons \cite{Wilkinson:2013kia,Boehm:2014vja,Schewtschenko:2015rno,Ali-Haimoud:2015pwa,Stadler:2018jin,Escudero:2018thh,Kumar:2018yhh,Diacoumis:2018ezi,Dey:2022ini,Moline:2019zev}, 
neutrinos \cite{Mangano:2006mp,Wilkinson:2014ksa,Escudero:2015yka,DiValentino:2017oaw,Olivares-DelCampo:2017feq,Diacoumis:2018ezi,Stadler:2019dii,Mosbech:2020ahp,Paul:2021ewd,Hooper:2021rjc,Mosbech:2022uud,Brax:2023tvn} 
or baryons \cite{Chen:2002yh,Melchiorri:2007sq,Ali-Haimoud:2015pwa,Dvorkin:2013cea,Munoz:2015bca,Kadota:2016tqq,Munoz:2017qpy,Gluscevic:2017ywp,Boddy:2018kfv,Xu:2018efh,Barkana:2018lgd,Slatyer:2018aqg,Munoz:2018jwq,Boddy:2018wzy,Nadler:2019zrb,DES:2020fxi,Maamari:2020aqz,Rogers:2021byl}) 
or dark radiation \cite{Diamanti:2012tg,Cyr-Racine:2013fsa,Chu:2014lja,Buen-Abad:2015ova,Lesgourgues:2015wza,Cyr-Racine:2015ihg,Foot:2016wvj,Krall:2017xcw,Archidiacono:2017slj,Buen-Abad:2017gxg,Archidiacono:2019wdp,Sameie:2018juk,Bohr:2021bdm,Schaeffer:2021qwm}
induce damping of the primordial density fluctuations via dark acoustic oscillations (DAO) in the early universe. This collisional damping suppresses structure formation on small scales \cite{Boehm:2000gq,Boehm:2004th}, which can potentially solve the small-scale problems \cite{Boehm:2014vja,Schewtschenko:2015rno,Vogelsberger:2015gpr}.
This leaves characteristic signatures on the cosmic microwave background (CMB) as well as the large-scale structure (LSS). In addition, the resulting suppression reduces the number of DM subhalos in a larger host halo and then observable satellite galaxies, which can be formed within DM subhalos but probably not in all subhalos \cite{Brooks:2012vi,Sawala:2014hqa}.

In this work, we study the constraints on the elastic scattering cross sections of DM and neutrinos using the latest observational data of Milky-Way satellite galaxies from the Dark Energy Survey (DES) and PanSTARRS1 (PS1) \cite{DES:2019vzn,DES:2019ltu}.\footnote{\href{https://github.com/eonadler/subhalo_satellite_connection}{https://github.com/eonadler/subhalo\_satellite\_connection}}
To quickly derive theoretical prediction of the number of Milky-Way satellites, we study and use a modified version of a semi-analytical subhalo model based on the extended Press-Schechter (EPS) formalism \cite{Press:1973iz,Bond:1990iw,Bower:1991kf,Lacey:1993iv} and subhalos' tidal evolution prescription, as developed in refs.~\cite{Hiroshima:2018kfv,Dekker:2021scf}.
The primordial density fluctuations collapse gravitationally, forming DM halos. 
Halos grow by accreting smaller halos, and the accreted halos are called subhalos.
After the accretion of subhalos onto a host halo, they lose mass through gravitational tidal stripping, changing their internal structure \cite{Jiang:2014nsa,vandenBosch:2004zs,Giocoli:2007uv}. 
The model developed in refs.~\cite{Hiroshima:2018kfv,Dekker:2021scf} describes the complementary semi-analytical evolution of a host halo and subhalos, including tidal stripping, with analytical expressions for the hierarchical assembly of DM halos provided by the EPS formalism.
In this study, we extend it to interacting dark matter models by modifying the EPS formalism, in particular the window function and the critical overdensity.
To obtain the expected number of subhalos, we use a modified version of publicly available \lib{SASHIMI} codes.\footnote{\href{https://github.com/shinichiroando/sashimi-c}{https://github.com/shinichiroando/sashimi-c}} \footnote{\href{https://github.com/shinichiroando/sashimi-w}{https://github.com/shinichiroando/sashimi-w}}

We compare the computed number of Milky-Way satellite galaxies with the observed number of satellite galaxies to constrain the cross section of DM and neutrinos. We use observational data of 270 Milky-Way satellite galaxies from DES and PS1 after completeness correction \cite{DES:2019vzn} as well as a subset of 94 satellite galaxies that contain kinematics data.
We consider the mass of the Milky-Way halo in the range of $(0.6\text{--}2.0)\times 10^{12}M_\odot$, which is determined in refs.~\cite{Karukes:2019jwa,Posti_2019,Eadie_2019,Fritz_2018} based on the latest Gaia data. For our conservative constraints, we assume that all the subhalos host satellite galaxies. 
We also derive stronger limits, imposing a galaxy forming condition in subhalos that reduces the expected number of satellite galaxies.

We find the strongest constraints on the DM-neutrino scattering cross section of $\sigma_{{\rm DM}\text{--}\nu,n}\propto E_\nu^n$ $(n=0,2,4)$ at $95\%$ confidence level (CL) in figures~\ref{fig:Const_DM_Nu_0}, \ref{fig:Const_DM_Nu_2} and \ref{fig:Const_DM_Nu_4}, where $E_\nu$ is the neutrino energy. 
These results are independent of specific particle physics model. 
The constraints from CMB \cite{Mangano:2006mp,Wilkinson:2014ksa,Escudero:2015yka,Diacoumis:2018ezi,Mosbech:2020ahp} and Lyman-$\alpha$ forests \cite{Wilkinson:2014ksa,Hooper:2021rjc} are also shown in these figures for comparison.
To compare with astrophysical constraints from observations of supernova (SN) 1987A neutrinos \cite{Mangano:2006mp} and high-energy neutrinos \cite{Arguelles:2017atb,Pandey:2018wvh,Alvey:2019jzx,Choi:2019ixb,Murase:2019xqi,Cline:2022qld,Ferrer:2022kei,Cline:2023tkp}, we need to consider a more specific model rather than the cross sections as discussed in section~\ref{sec6.2}.

This paper is organized as follows. In section~\ref{sec2}, we review DM-neutrino interactions and their effects on primordial density fluctuations. 
Section~\ref{sec3} presents our semi-analytical model of subhalo evolution for the DM-radiation interaction scenarios. 
In section~\ref{sec4}, we compare the results of our model with publicly available data of N-body simulations for interacting DM model.
In section~\ref{sec5}, we show the constraints on the DM-neutrinos cross sections using the presented semi-analytical model and the population data of Milky-Way satellites.
In section~\ref{sec6}, we discuss uncertainties of our results and compare our constraints with astrophysical ones on the DM-neutrino scattering. Here we also compare our results with the results in the literature.
Section~\ref{sec7} concludes the paper.
Supplemental materials on our semi-analytical subhalo model are given in appendices~\ref{appa}, \ref{appb} and \ref{appc}.

\section{DM-neutrino interactions}
\label{sec2}

We consider three types of the scattering cross section of DM and neutrinos, $\sigma_{{\rm DM}\text{--}\nu}$. One is constant and the others are proportional to neutrino-energy squared and quartered, $\sigma_{{\rm DM}\text{--}\nu,n}\propto E_\nu^n \propto a^{-n}$ $(n=0,2,4)$, where $a$ is the scale factor of the universe normalized to unity today. 
A constant cross section can be obtained when the DM and mediator masses are strongly degenerate (like Thomson scattering). The cross section would typically depend on neutrino-energy squared or quartered (see appendix~C in ref.~\cite{Arguelles:2017atb} for the cross sections in specific DM-neutrino interactions).

DM-radiation interactions result in additional collision terms in the perturbation equations for DM and radiation.
The modified Euler equations for DM and neutrinos are given by\footnote{We use the Newtonian gauge and assume massless neutrinos.}
\cite{Boehm:2001hm,Mangano:2006mp,Wilkinson:2014ksa}
\begin{align}
\dot{\theta}_{\rm \nu}&= k^2\psi+k^2\left(\frac{1}{4}\delta_\nu-\sigma_\nu \right)-\Gamma_{\nu\text{--}{\rm DM}}(\theta_\nu-\theta_{\rm DM}), \\
\dot{\theta}_{\rm DM}&=k^2\psi-\mathcal{H}\theta_{\rm DM}-\Gamma_{{\rm DM\text{--}}\nu}(\theta_{\rm DM}-\theta_\nu),
\end{align}
where $\theta_\nu$ and $\theta_{\rm DM}$ are the velocity divergence for the neutrinos and DM, respectively, $k$ is the comoving wavenumber, $\psi$ is the curvature potential, $\delta_\nu$ is the neutrino density fluctuation, $\sigma_\nu$ is the neutrino anisotropic stress potential and $\mathcal{H}=(\dot{a}/a)$ is the conformal Hubble parameter. An overdot denotes a derivative with respect to conformal time. The DM-neutrino interactions also modify the Boltzmann hierarchy for neutrinos. For brevity, we do not show the full modified Boltzmann equations for DM and neutrinos \cite{Stadler:2019dii,Mosbech:2020ahp}.

$\Gamma_{\nu\text{--}{\rm DM}}$ ($\Gamma_{{\rm DM\text{--}}\nu}$) is the conformal neutrino (DM) collision rate with DM (neutrinos),
\begin{align}
\Gamma_{\nu\text{--}{\rm DM}}&= a\sigma_{{\rm DM\text{--}}\nu}n_{\rm DM}, \label{NuIR} \\
\Gamma_{{\rm DM\text{--}}\nu}&=\frac{4\rho_\nu}{3\rho_{\rm DM}}\Gamma_{\nu\text{--}{\rm DM}},
\label{DMIR}
\end{align}
where $n_{\rm DM}=\rho_{\rm DM}/m_{\rm DM}$ is the DM number density and $m_{\rm DM}$ is the DM mass.
The unknown parameter of DM in eqs.~(\ref{NuIR}) and (\ref{DMIR}) is $\sigma_{{\rm DM\text{--}}\nu}/m_{\rm DM}$, which can be written as the dimensionless quantity,
\begin{align}
u_{{\rm DM\text{--}}\nu,n}&\equiv\left[\frac{\sigma_{{\rm DM\text{--}}\nu,n}}{\sigma_{\rm Th}} \right]\left[\frac{m_{\rm DM}}{100\ {\rm GeV}} \right]^{-1}, \label{udef} \\
u_{{\rm DM\text{--}}\nu,n} &= u_{{\rm DM\text{--}}\nu,n}^0a^{-n}, \label{udef2}
\end{align}
where $\sigma_{\rm Th}=6.65\times 10^{-25}\ {\rm cm^2}$ is the Thomson cross section and $u_{{\rm DM\text{--}}\nu,n}^0$ is a value of $u_{{\rm DM\text{--}}\nu,n}$ normalized by relic neutrino momentum today.

\begin{figure}
    \centering
    \includegraphics[width = 10cm]{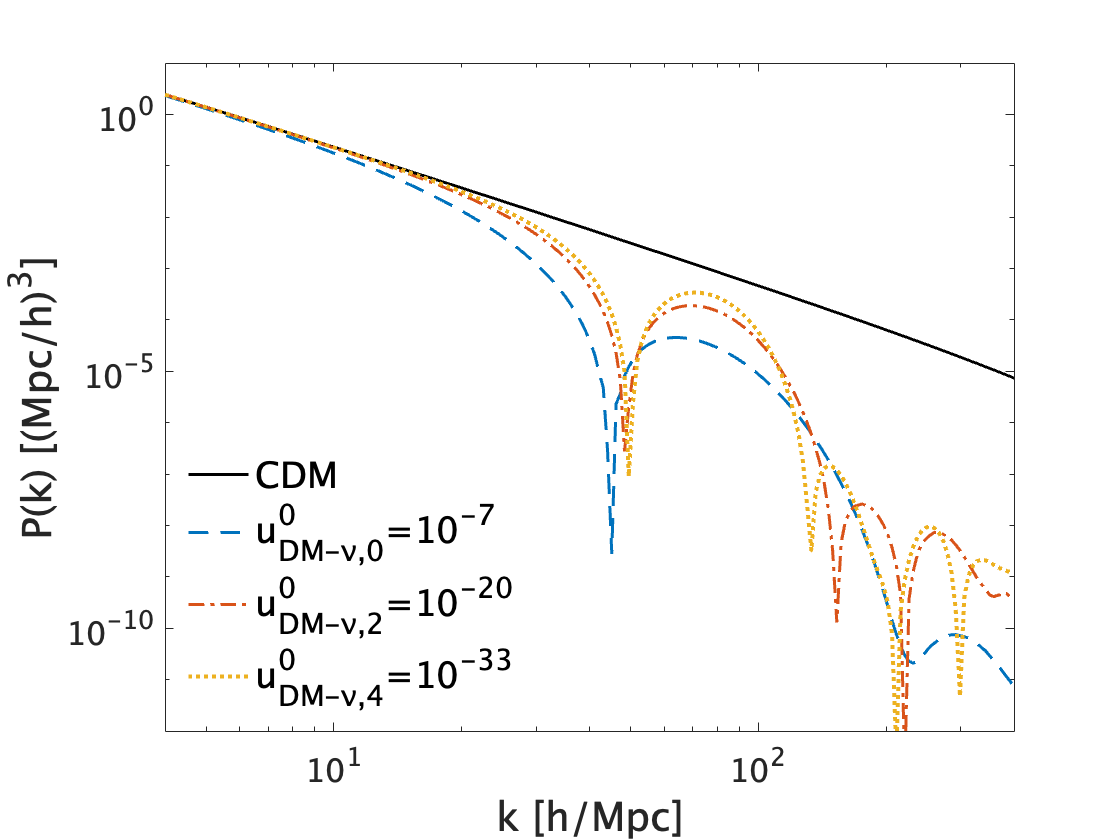}
    \caption{The linear matter power spectra for CDM, the DM-neutrino interaction scenario with $u_{\rm DM-\nu,0}^0=10^{-7}$, $u_{\rm DM-\nu,2}^0=10^{-20}$ and $u_{\rm DM-\nu,4}^0=10^{-33}$, which are defined in eqs.~(\ref{udef}) and (\ref{udef2}). 
    }
    \label{fig:Pk_radiation}
\end{figure}

In figure~\ref{fig:Pk_radiation}, we show the effects of DM-neutrino interactions on the linear matter power spectrum.
We obtain the matter power spectrum using publicly available modified versions of the Boltzmann code \lib{CLASS} \cite{Blas:2011rf} described in refs.~\cite{Stadler:2019dii,Becker:2020hzj,Mosbech:2020ahp} for DM-neutrino interactions.
For higher $n$, the decoupling time scale is more instantaneous, inducing the higher frequency of oscillations and the flatter spectrum shape.

For the interaction strength of our interest, DM decouples with neutrinos long before recombination \cite{Becker:2020hzj}. We expect that the DM-neutrino interactions affect only the initial matter power spectrum and have negligible direct effects on the late-time evolution of DM.

Note that we assume the DM energy density behaves as $\rho_{\rm DM}=\rho_{\rm DM}^0a^{-3}$ at the DM-neutrino decoupling epoch, where $\rho_{\rm DM}^0$ is the DM energy density observed today. As a result, for $m_{\rm DM}\lesssim 10\ {\rm keV}$ with $u_{\rm DM\text{--}\nu}$ of our interest, DM may be enough heated by neutrinos to be relativistic at the DM-neutrino decoupling \cite{Becker:2020hzj} (see also eq.~(\ref{zdec}) in appendix~\ref{appa}), changing the evolution of $\rho_{\rm DM}$. We focus on the DM mass of $m_{\rm DM}\gtrsim 10\ {\rm keV}$.

\section{Semi-analytical subhalo model for dark acoustic oscillations}
\label{sec3}

To quickly and comprehensively estimate the effects of the suppressed matter power spectrum due to DM-radiation interactions on the number of subhalos via DAO, we consider a modified version of semi-analytical model for subhalo (and host halo) evolution proposed in refs.~\cite{Hiroshima:2018kfv,Dekker:2021scf}. 
Our semi-analytical model is similar to the case of WDM particles \cite{Dekker:2021scf}, where the small scale structure is similarly suppressed due to their free streaming, but we modify the EPS formalism, in particular the window function and the critical overdensity, as described in sections~\ref{sec3.1} and \ref{sec3.2}.

\subsection{Variance of density fluctuations and window function}
\label{sec3.1}

An important ingredient in the semi-analytical model based on the EPS formalism is the linear matter power spectrum characterized by the variance, $S=\sigma^2$ at $z=0$, where $z$ is the redshift. In the EPS method, the variance has to be spherically smoothed on a mass scale $M$ weighted by a window function $\widetilde{W}(kR)$,
\begin{align}
S(M)=\sigma^2(M)=\frac{1}{2\pi^2}\int_0^{\infty} P(k) \widetilde{W}^2(kR)k^2 dk,
\end{align}
where $M=M(R)$.
$\widetilde{W}(kR)$ needs to be fixed by comparing with N-body simulations. The top-hat filter is a successful choice in the CDM case, while 
the sharp-k filter successfully reproduces the prediction with N-body simulations for the truncated power spectra such as WDM \cite{Benson:2012su,Lovell:2015psz}. 
However, both window functions fail to reproduce the halo abundance in the universe for interacting DM models \cite{Schewtschenko:2014fca,Sameie:2018juk} since they cannot properly account for the suppressed power spectrum on small scale.
We adopt another window function, called the smooth-k filter proposed in ref.~\cite{Leo:2018odn},
\begin{align}
\widetilde{W}^{\rm smooth-k}(kR)=\frac{1}{1+(kR)^\beta},
\end{align}
where $\beta$ is a free parameter that controls the small-scale slope. For $\beta\rightarrow \infty$, the smooth-k filter coincides with the sharp-k filter, $\widetilde{W}^{\rm smooth-k}(kR)\rightarrow \widetilde{W}^{\rm sharp-k}(kR)=\Theta(1-kR)$, where $\Theta$ is the Heaviside step function.
For large $k$ modes, the smooth-k filter has non-zero values, unlike the sharp-k filter. It successfully reproduces the halo abundance in some models of interacting DM with dark radiation (DR) \cite{Sameie:2018juk,Bohr:2021bdm,Schaeffer:2021qwm}, where the matter power spectrum is similar to figure~\ref{fig:Pk_radiation}.

The mass $M$ associated with the filter scale $R$ is less well defined. We assign the mass $M$ as $M=4\pi\bar{\rho}(cR)^3/3$, where $\bar{\rho}$ is the average matter density of the universe and $c$ is a free parameter. The parameter set $(c,\beta)$ must be calibrated using simulations. We adopt $(c,\beta)= (3.7, 3.5)$ \cite{ Sameie:2018juk,Bohr:2021bdm}. Similar but different choices of $(c,\beta)$ for interacting DM are also proposed in refs.~\cite{Leo:2018odn,Schaeffer:2021qwm}.

\subsection{Critical overdensity for collapse}
\label{sec3.2}
Another important ingredient in the EPS formalism is the critical overdensity $\delta_c$, above which
the DM linear density spherically collapses into a halo.
In the DM-radiation interaction scenario, radiation may heat DM, producing pressure for DM and delaying the collapse. 
This may also change the critical overdensity from the CDM case and be similar to the WDM case. However, for $m_{\rm DM}\gtrsim {\rm MeV}$ with the interaction strength of our interest, DM would be heavy enough that we adopt the CDM case to $\delta_c$ (see also the detailed discussion in appendix~\ref{appa}),
\begin{align}
\delta_c \simeq \delta_{c,{\rm CDM}}(z)\equiv1.686/D(z),
\label{CDM_oc}
\end{align}
where $D(z)$ is the linear growth factor normalized to unity at $z=0$.
For $m_{\rm DM}\lesssim {\rm MeV}$, using eq.~(\ref{CDM_oc}) is still a conservative choice.

\subsection{Evolution of subhalo structure}
\label{sec3.3}

After the accretion of subhalos onto the host halo, they lose mass by tidal stripping in the host halo.
The average mass-loss rate of subhalos is given by \cite{Jiang:2014nsa}
\begin{align}
\dot{m}(z)=-A\frac{m(z)}{\tau_{\rm dyn}(z)}\left[\frac{m(z)}{M(z)} \right]^\zeta,
\label{mevol}
\end{align}
where $m(z)$ and $M(z)$ are the subhalo and host halo masses, respectively, and $\tau_{\rm dyn}(z)$ is the dynamical timescale \cite{Jiang:2014nsa}.
$A$ and $\zeta$ are functions of $M(z)$ and $z$.
The mass evolution of the host halo is the same in the CDM and WDM universes \cite{Hiroshima:2018kfv,Dekker:2021scf} because the accretion of larger halos may mainly contribute to the evolution of the host halo. The functions $A$ and $\zeta$ are also almost the same in the CDM and WDM universes \cite{Hiroshima:2018kfv,Dekker:2021scf} because the mass evolution of subhalos would mainly depend on the host halo mass.
We expect that even for interacting DM models, the subhalo mass evolution is the same with the CDM and WDM cases.
We adopt the analytical expression of the host halo evolution $M(z)$ in ref.~\cite{Correa:2014xma},
\begin{align}
M(z)=M_0(1+z)^\alpha\exp(\beta z),
\label{Mevol}
\end{align}
where
\begin{align}
\beta=-f(M_0),\ \ \ \ \alpha=\left[1.686(2/\pi)^{1/2}\frac{dD}{dz}\biggl|_{z=0}+1\right]f(M_0),
\end{align}
with
\begin{align}
&f(M_0)=[S(M_0/q)-S(M_0)]^{-1/2}, \nonumber \\
&q=4.137\tilde{z}_f^{-0.9476}, \nonumber \\
&\tilde{z}_f=-0.0064(\log_{10}M_0)^2+0.0237(\log_{10}M_0)+1.8827.
\end{align}
$M(z)$ is further generalized to obtain the host halo mass $M(z_i)$ at redshift $z_i$ as \cite{Correa:2015dva}
\begin{align}
M(z)=M(z_i)(1+z-z_i)^\alpha\exp\left(\beta(z-z_i)\right).
\end{align}
We also adopt the fitting functions of $A$ and $\zeta$ proposed in ref.~\cite{Dekker:2021scf},
\begin{align}
&\log A=\left[-0.0019\log\left(\frac{M(z)}{M_\odot} \right)+0.045 \right]z+0.0097\log\left(\frac{M(z)}{M_\odot} \right)-0.31, \\
&\zeta=\left[-0.000056\log\left(\frac{M(z)}{M_\odot} \right)+0.0014 \right]z+0.00033\log\left(\frac{M(z)}{M_\odot}\right)-0.0081.
\end{align}

We assume the subhalo density profile follows the Navarro-Frenk-White (NFW) profile \cite{Navarro:1996gj} with a truncation radius $r_t$ due to tidal stripping \cite{Springel:2008cc},
\begin{align}
\rho(r)=
\left\{
\begin{array}{ll}
\rho_sr_s^3/[r(r+r_s)^2], & {\rm for}\ \ r \leq r_t, \\
0, & {\rm for}\ \ r > r_t.
\end{array}
\right.
\end{align}
We also assume the NFW density profile for (sub)halos before their accretion onto a host halo. 

At accretion redshift $z_a$ when the halo becomes a subhalo, we determine the virial radius $r_{{\rm vir},a}$ for a subhalo with mass $m_a$ as
\begin{align}
m_a=\frac{4\pi}{3}\Delta(z_a)\rho_c(z_a)r_{{\rm vir},a}^3,
\end{align}
where $\Delta=18\pi^2+82d-39d^2$, $d(z)=\Omega_m(1+z)^3/[\Omega_m(1+z)^3+\Omega_\Lambda]-1$ \cite{Bryan:1997dn} and $\rho_c(z)$ is the critical density.
The scale radius $r_{s,a}$ at $z_a$ is determined as $r_{s,a}=r_{{\rm vir},a}/c_{{\rm vir},a}$, given a virial concentration parameter $c_{{\rm vir},a}$.
Then the characteristic density $\rho_{s,a}$ at $z_a$ is obtained by 
\begin{align}
\rho_{s,a}=\frac{m_a}{4\pi r_{s,a}^3f(c_{\rm vir,a})},
\end{align}
where $f(x)=\ln (1+x)-x/(1+x)$. 
The suppression of power spectrum on small scales causes DM halos to form later than in the CDM case. The concentration of halos would be lower than the CDM case due to a lower background density owning to cosmic expansion. 
We adopt the mean values of the concentration parameter obtained in ref.~\cite{Ludlow:2016ifl}. In ref.~\cite{Ludlow:2016ifl}, the formula for the concentration is applied to CDM and WDM. Recently, ref.~\cite{Bohr:2021bdm} confirms that this formula is in a good agreement with simulations in the DM-DR interaction case.

The parameters $r_s$ and $\rho_s$ are related to the maximum circular velocity $V_{\rm max}$ and the corresponding radius $r_{\rm max}$ as
\begin{align}
r_s=\frac{r_{\rm max}}{2.163},\ \ \ \ 
\rho_s=\frac{4.625}{4\pi G}\left(\frac{V_{\rm max}}{r_s} \right)^2.
\label{rsrhos_rel}
\end{align}
The subhalo structure before and after tidal stripping is related to $V_{\rm max}$ and $r_{\rm max}$ at accretion redshift $z_a$ and any later redshift $z_0$ \cite{Hiroshima:2018kfv,Pe_arrubia_2010},
\begin{align}
\frac{V_{{\rm max},0}}{V_{{\rm max},a}}=\frac{2^{0.4}(m_0/m_a)^{0.3}}{(1+m_0/m_a)^{0.4}},\ \ \ \ 
\frac{r_{\rm max,0}}{r_{{\rm max},a}}=\frac{2^{-0.3}(m_0/m_a)^{0.4}}{(1+m_0/m_a)^{-0.3}},
\end{align}
where $m_0$ is the subhalo mass at $z_0$ after tidal stripping. We obtain $r_{s,0}$ and $\rho_{s,0}$ at $z_0$ through eqs.~(\ref{rsrhos_rel}).
Then the truncation radius $r_{t,0}$ is obtained by solving 
\begin{align}
m_0=4\pi\rho_{s,0}r_{s,0}^3f\left(\frac{r_{t,0}}{r_{s,0}} \right)
\end{align}
and eq.~(\ref{mevol}). 
When $r_{t,0}/r_{s,0}<0.77$, subhalos may be completely disrupted \cite{Hayashi:2002qv}, but it has also been pointed out that its complete disruption may not occur \cite{vandenBosch:2017ynq}.
In this work, we take into account its complete tidal disruption.

\subsection{Evolution of the number of subhalos}
\label{sec3.4}

The number of subhalos as a function of mass can be computed as
\begin{align}
\frac{dN_{\rm sh}}{dm}&=\int dm_a \int dz_a \frac{d^2N_a}{dm_adz_a}\int dc_a P(c_a|m_a,z_a) \nonumber \\
&\ \ \ \ \times \delta(m-m(m_a,z_a,c_a))
\Theta\left(r_t(m_a,z_a,c_a)-0.77r_s(m_a,z_a,c_a)\right),
\label{dNdm}
\end{align}
where $\delta(x)$ is the Dirac delta function, $c_a$ are the concentration parameter at accretion redshift $z_a$ and $P(c_a|m_a,z_a)$ is the distribution of $c_a$.
We can calculate the number of subhalos as a function of other quantities such as $r_s,\ \rho_s$ and $V_{\rm max}$ by replacing the argument of the delta function in eq.~(\ref{dNdm}) appropriately.
For $P(c_a|m_a,z_a)$, we adopt the log-normal function with the mean value $\bar{c}(m_a,z_a)$ \cite{Ludlow:2016ifl} and standard deviation of $\sigma_{\log c}=0.13$ \cite{Ishiyama:2011af}.

The number of subhalos accreted onto the main branch of host halo with mass $m_a$ at accretion redshift $z_a$, $d^2N_a/dm_adz_a$, can be computed based on the EPS formalism as \cite{Yang:2011rf}
\begin{align}
 \frac{d^2N_{\rm a}}{d m_adz_a}=\frac{1}{m_a}\mathcal{F}(s_a,\delta_a|S_0,\delta_0;\overline{M}_a)\frac{ds_a}{dm_a}\frac{d\overline{M}_a}{dz_a},
\end{align}
where $\overline{M}_a(z_a)$ is the mean mass of the main branch of host halo at $z_a$ given by eq.~(\ref{Mevol}). 
$s_a\equiv \sigma^2(m_a)$ and $S_0\equiv \sigma^2(M_0)$ are the variances of liner matter density on mass scales of $m_a$ and $M_0$, respectively. $\delta_a\equiv \delta_c(z_a)$ and $\delta_0\equiv \delta_c(z_0)$ are the critical overdensities.
$\mathcal{F}(s_a,\delta_a|S_0,\delta_0;\overline{M}_a)$ is the fraction of accreted subhalos with mass $m_a$ at $z_a$ in the mass range of the main branch of the host halo $[\overline{M}_a-d\overline{M}_a,\ \overline{M}_a]$ defined as
\begin{align}
    \mathcal{F}(s_a,\delta_a|S_0,\delta_0;\overline{M}_a)=\int \Phi(s_a,\delta_a|S_0,\delta_0; M_a)P(M_a|S_0,\delta_0)dM_a,
\end{align}
with
\begin{align}
&\Phi=\left[\int^{\infty}_{S(m_{\rm max})}F(s_a,\delta_a|S_0,\delta_0;M_a) ds_a \right]^{-1}\times
\left\{
\begin{array}{ll}
F(s_a,\delta_a|S_0,\delta_0;M_a) & \text{if}\ m_a \geq m_{\rm max}  \\
0 & \text{if}\ m_a < m_{\rm max}
\end{array}
\right., \\
&F(s_a,\delta_a|S_0,\delta_0;M_a)d\ln \delta_a=\frac{1}{\sqrt{2\pi}}\frac{\delta_a-\delta_M}{(s_a-S_M)^{3/2}}\exp\left[-\frac{(\delta_a-\delta_M)^2}{2(s_a-S_M)} \right],
\end{align}
where $P(M_a|S_0,\delta_0)$ is the distribution of the main branch of host halo $M_a$ and $[...]^{-1}$ is the normalization factor.
$m_{\rm max}\equiv \min[M_a,\ M_0/2]$ is the maximum mass of subhalos accreting onto the main branch and $M=M_{\rm max}\equiv \min[M_a+m_{\rm max},\ M_0]$ is the maximum mass of the main branch of host halo after the mass of the main branch has increased due to the accretion of subhalos with $m_a$. $\delta_M$ are defined at the redshift when the main branch has mass $M_{\rm max}$ and $S_M=\sigma^2(M_{\rm max})$.


\section{Comparison with N-body simulations}
\label{sec4}

We test our semi-analytical model with numerical simulations.
Unfortunately, there is no numerical simulation of subhalo evolution for interacting DM models with neutrinos in the literature. However, there are some results of N-body simulations for subhalo population for interacting DM models with DR \cite{Vogelsberger:2015gpr}.
We compare the results from our model with these numerical results.
Any radiation would also affect only the initial matter power spectrum directly. 

\begin{figure}
    \centering
    \includegraphics[width = 10cm]{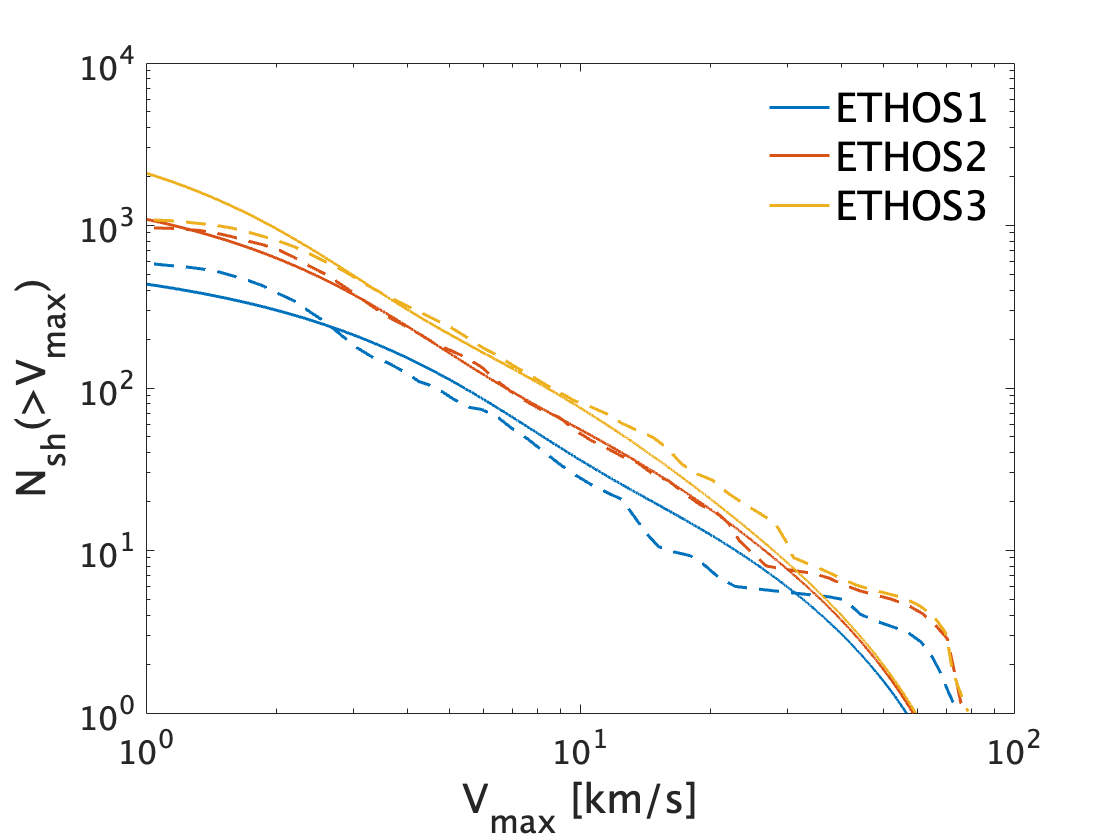}
    \caption{Cumulative number of subhalos as a function of the maximum circular velocity with the host halo mass of $M_{200}=1.6\times 10^{12}M_\odot$ for DM-DR interaction scenarios (ETHOS1-ETHOS3 models described in ref.~\cite{Vogelsberger:2015gpr}). Solid lines denote our results and dashed lines denote the results of N-body simulations from  ref.~\cite{Vogelsberger:2015gpr}.}
    \label{fig:ETHOS_subhalo}
\end{figure}

The DM-DR interactions are parametrized by the DM and mediator masses, the abundance of DR and the DM-mediator and DR-mediator couplings.
In refs.~\cite{Vogelsberger:2015gpr,Cyr-Racine:2015ihg}, the effective theory of structure formulation (ETHOS) including DM-DR interactions and DM self-interaction is proposed\footnote{DM self-interactions would not affect the number of subhalos significantly \cite{Vogelsberger:2015gpr}.}. In the following, we focus on ETHOS1-ETHOS3 models, where the above parameters are fixed to concrete values (see ref.~\cite{Vogelsberger:2015gpr} for details).
In figure~\ref{fig:ETHOS_subhalo}, we compare the cumulative number of subhalos as a function of maximum circular velocity with the numerical results for the host halo mass of $M_{\rm 200}=1.6\times 10^{12}M_\odot$.
We find our model to be in very good agreement with the N-body simulations within a factor of 1.8.

We should note that there are also some results of N-body simulations for the DM-photon interaction scenarios \cite{Boehm:2014vja,Moline:2019zev}. In ref.~\cite{Boehm:2014vja}, the authors compute the number of subhalos as a function of the maximum circular velocity $V_{\rm max}$ in the mass bin of $(2.3\text{--}2.7)\times 10^{12}M_\odot$ but they calculate $V_{\rm max}$ by the observed stellar line-of-sight velocity dispersions, $V_{\rm max}=\sqrt{3}\sigma$, which may be underestimated.
In ref.~\cite{Moline:2019zev}, the authors compute the number of subhalos as a function of subhalo mass in the large mass bin of $(0.3\text{--}1.4)\times 10^{12}M_\odot/h$. Some details of precision of the simulations in refs.~\cite{Boehm:2014vja,Moline:2019zev} are also lower than in the simulations in the case of ETHOS models \cite{Vogelsberger:2015gpr}.
Thus, since the simulations in refs.~\cite{Boehm:2014vja,Moline:2019zev} might include relatively large uncertainties, we will not compare our results with them.

Our semi-analytical model is in a very good agreement with the ETHOS models with DM-DR interactions within a factor of 1.8.
In addition, our models with the smooth-k filter of different parameter values $(c,\beta)$ \cite{Leo:2018odn,Schaeffer:2021qwm} also predict very similar results of figure~\ref{fig:ETHOS_subhalo} as discussed in appendix~\ref{appb}. 
These results indicate our model is a valuable tool to explore DM substructure in various interacting DM models quickly.
Note that the parameter set of $(c,\beta)$ in the smooth-k filter is determined to account for the power spectrum on small scale properly. As in figure~\ref{fig:Pk_radiation}, the small-scale shape of the power spectrum is flatter for higher $n$ of radiation-energy dependence. This indicates that for better predictions of subhalo properties, 
the parameter set of $(c,\beta)$ might slightly change, depending on the nature of interactions (interacting species, the radiation-energy dependence of the cross section, the DR abundance, etc).

\section{Constraints on DM-neutrino scattering}
\label{sec5}


\subsection{Constraints on DM-neutrino scattering cross sections}
\label{sec5.1}

We compare the expected number of subhalos in the Milky Way obtained by integrating eq.~(\ref{dNdm}) with the observed number of Milky-Way satellite galaxies from DES and PS1 reported in ref.~\cite{DES:2019vzn}. Ref.~\cite{DES:2019vzn} corrects the number of detected satellites in DES and PS1 by volumic weights relative to the observable volume. This completeness correction predicts the Milky Way contains $270$ satellite galaxies within $300$ kpc with absolute $V$-band magnitude of $M_V<0$. 

We also consider the stellar kinematics data of the Milky-Way satellites. The observed satellite with the smallest line-of-sight velocity dispersion is Leo V with $\sigma=2.3\ {\rm km\ s^{-1}}$ \cite{Simon:2019nxf}.
Assuming that the velocity dispersion is isotropic, the halo circular velocity of Leo V is $V_{\rm circ}=\sqrt{3}\sigma=4\ {\rm km\  s^{-1}}$. 
We conservatively regard $V_{\rm circ}=4\ {\rm km\  s^{-1}}$ as the threshold of maximum circular velocity.
In ref.~\cite{Dekker:2021scf}, the authors found 94 satellite galaxies with $V_{\rm max}>4\ {\rm km\ s^{-1}}$, which contains 82 satellite galaxies after completeness corrections and 12 satellite galaxies without the corrections.

We compute eq.~(\ref{dNdm}) taking 500 logarithmic steps from $m_a=10M_\odot$ to $0.1M_{200}$ in the subhalo mass and steps of $dz=0.1$ from $z_a=7$ to $0.1$ in redshift, where $M_{200}$ is the Milky-Way halo mass. $M_{200}$ is the mass contained within the radius whose mean density 200 times the critical density. We have confirmed that the results converge enough in these steps. 
We consider the Milky-Way mass in the range of $M_{200}=(0.6\text{--}2.0)\times 10^{12}M_\odot$ based on the latest Gaia data \cite{Karukes:2019jwa,Posti_2019,Eadie_2019,Fritz_2018}.
To obtain the expected number of subhalos, we modify the publicly available \lib{SASHIMI} code.

To constrain the elastic scattering between DM and neutrinos, we define the Poisson probability of obtaining the observed number of satellites $N$ for the expected number of satellites in our models $\mu$ as $P(N|\mu)=\mu^N\exp(-\mu)/N!$. We rule out the elastic scattering of DM and neutrinos at 95\% CL when $P(>N_{\rm obs}|\mu)=\sum_{N=N_{\rm obs}}^{N=\infty}P(N|\mu)<0.05$.
For the observational data of satellite galaxies of $N_{\rm obs}=270$, $\mu\leq243$ is ruled out while for the kinematics data of $N_{\rm obs}=94$ with $V_{\rm max}>4\ {\rm km\ s^{-1}}$, $\mu\leq78$ is ruled out.

\begin{figure}
    \centering
    \includegraphics[width = 10cm]{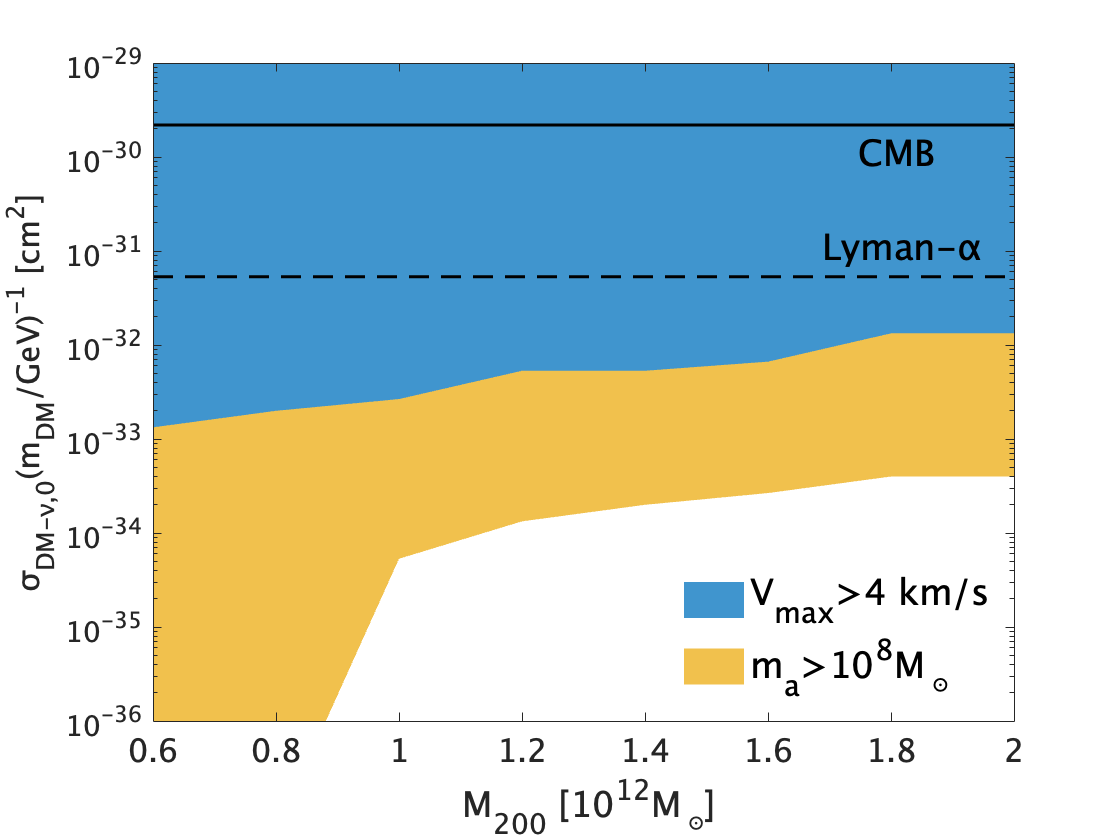}
    \caption{Constraints on the DM-neutrino cross section as $\sigma_{{\rm DM}\text{--}\nu,0}=$ const at 95 CL as a function of the Milky-Way mass considering the kinematics data of 94 Milky-Way satellites with $V_{\rm max}>4\ {\rm km/s}$ (blue) as well as the data of 270 Milky-Way satellites imposing the satellite forming condition of $m_a>10^8M_\odot$ (yellow). The constraints from CMB (solid line) \cite{Diacoumis:2018ezi,Mosbech:2020ahp} and Lyman-$\alpha$ (dashed line) \cite{Hooper:2021rjc} are shown for comparison.}
    \label{fig:Const_DM_Nu_0}
\end{figure}

\begin{figure}
    \centering
    \includegraphics[width = 10cm]{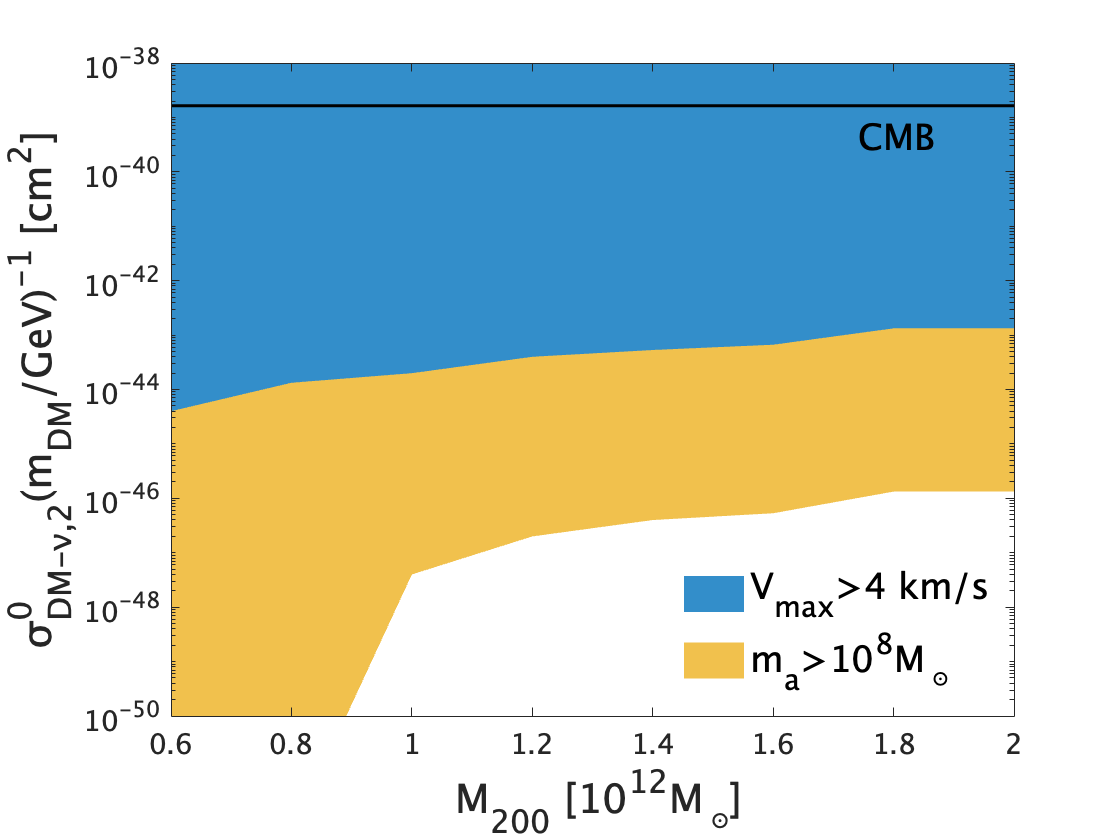}
    \caption{Constraints on the DM-neutrino cross section as $\sigma_{{\rm DM}\text{--}\nu,2}\propto E_\nu^2$ at 95 CL. $\sigma^0_{{\rm DM}\text{--}\nu,2}$ is the cross section normalized by relic neutrino momentum today, $\sigma_{{\rm DM}\text{--}\nu,2}=\sigma^0_{{\rm DM}\text{--}\nu,2}(E_\nu/E_\nu^0)^2$. The other details are the same as figure~\ref{fig:Const_DM_Nu_0}.}
    \label{fig:Const_DM_Nu_2}
\end{figure}

\begin{figure}
    \centering
    \includegraphics[width = 10cm]{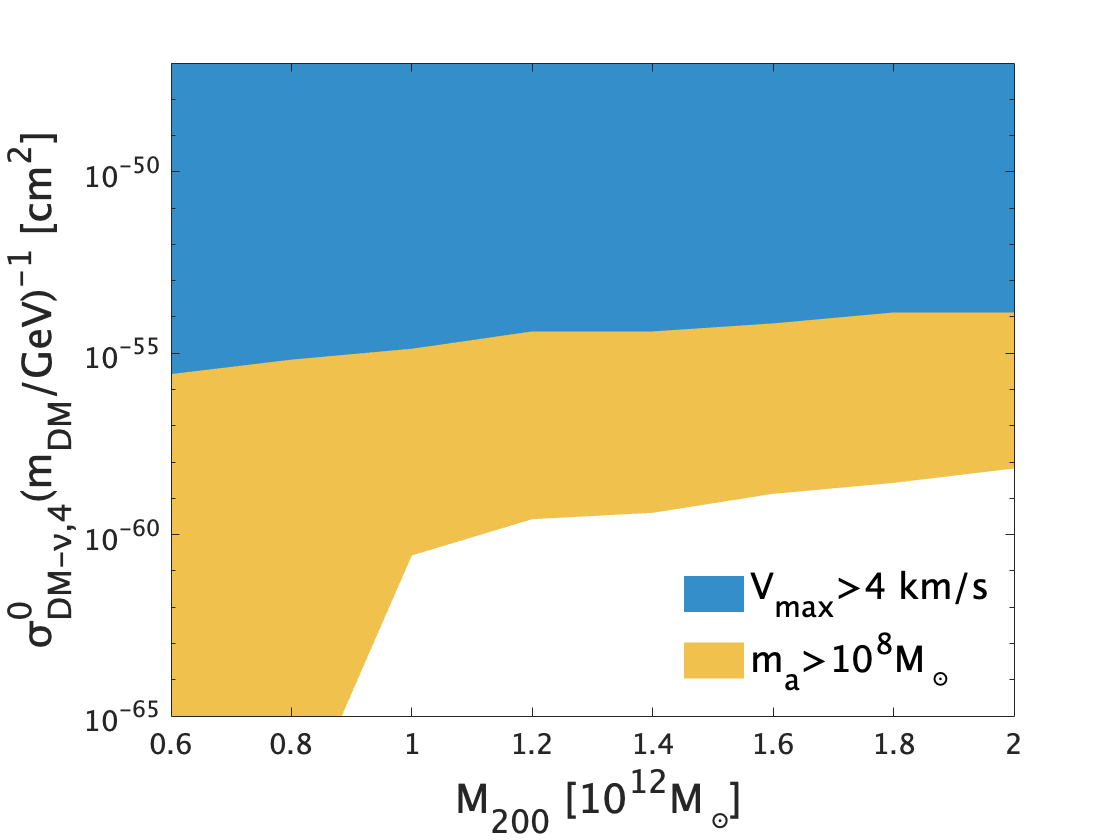}
    \caption{Constraints on the DM-neutrino cross section as $\sigma_{{\rm DM}\text{--}\nu,4}\propto E_\nu^4$ at 95 CL. $\sigma^0_{{\rm DM}\text{--}\nu,4}$ is the cross section normalized by relic neutrino momentum today, $\sigma_{{\rm DM}\text{--}\nu,4}=\sigma^0_{{\rm DM}\text{--}\nu,4}(E_\nu/E_\nu^0)^4$. The other details are the same as figure~\ref{fig:Const_DM_Nu_0}.}
    \label{fig:Const_DM_Nu_4}
\end{figure}

For our conservative constraints, we assume that all subhalos host satellite galaxies and use the kinematics data of 94 Milky-Way satellites with $V_{\rm max}>4\ {\rm km\ s^{-1}}$.
In figures~\ref{fig:Const_DM_Nu_0}, \ref{fig:Const_DM_Nu_2} and~\ref{fig:Const_DM_Nu_4}, we show the upper bounds on the DM-neutrino scattering cross section of $\sigma_{{\rm DM}\text{--}\nu,n}\propto E_\nu^n\ (n=0,2,4)$ at 95\% CL as a function of the Milky Way mass, respectively. 
We constrain the DM-neutrino cross section of $\sigma_{{\rm DM}\text{--}\nu,0}< 10^{-32}\ {\rm cm^2}\ (m_{\rm DM}/{\rm GeV})$, $\sigma_{{\rm DM}\text{--}\nu,2}< 10^{-43}\ {\rm cm^2}\ (m_{\rm DM}/{\rm GeV})(E_\nu/E_{\nu}^0)^2$ and $\sigma_{{\rm DM}\text{--}\nu,4}< 10^{-54}\ {\rm cm^2}\ (m_{\rm DM}/{\rm GeV})(E_\nu/E_{\nu}^0)^4$, where $E_\nu^0 \simeq 3.15 T_\nu^0 \simeq 6.1\ {\rm K}$.
$E_\nu^0$ and $T_\nu^0$ are the average momentum and temperature of relic cosmic neutrinos today, respectively.
These constraints are stronger than the constraints from CMB \cite{Diacoumis:2018ezi,Mosbech:2020ahp} and Lyman-$\alpha$ \cite{Hooper:2021rjc}, which are also shown in figures~\ref{fig:Const_DM_Nu_0} and \ref{fig:Const_DM_Nu_2}.

More realistically, not all subhalos would host satellite galaxies. 
An ultraviolet background produced by quasars and stars heats gas in small halos, allowing it to escape from their shallow gravitational potential and suppressing star formation in small halos.
Some satellite forming conditions reduce the expected number of satellite galaxies, imposing more stringent upper bounds on the scattering of DM and neutrinos. 
We adopt a threshold halo mass at accretion of $m_a\simeq 10^8M_\odot$ \cite{Schneider:2014rda,Sawala:2014hqa,Brooks:2012vi}, below which we assume that subhalos do not host satellite galaxies.
In this case we can impose the strongest upper bounds of $\sigma_{{\rm DM}\text{--}\nu,0}< 4\times 10^{-34}\ {\rm cm^2}\ (m_{\rm DM}/{\rm GeV})$, $\sigma_{{\rm DM}\text{--}\nu,2}< 10^{-46}\ {\rm cm^2}\ (m_{\rm DM}/{\rm GeV})(E_\nu/E_{\nu}^0)^2$ and $\sigma_{{\rm DM}\text{--}\nu,4}< 7\times 10^{-59}\ {\rm cm^2}\ (m_{\rm DM}/{\rm GeV})(E_\nu/E_{\nu}^0)^4$.

\subsection{Constraints on specific DM-neutrino interactions}
\label{sec5.2}

Following appendix~C in ref.~\cite{Arguelles:2017atb}, we map our strongest constraints to couplings of specific DM-neutrino interactions.\footnote{Neutrino telescopes can also constrain couplings of DM-neutrino interactions via DM annihilation to neutrinos \cite{Arguelles:2019ouk}. If DM does not annihilate today as in the scenario of asymmetric DM, these constraints are not imposed.}
We denote the DM and mediator masses as $m_{\rm DM}$ and $m_{\phi}$, respectively. We consider the case of $m_{\phi} \gtrsim m_{\rm DM} \gg E_\nu$.

\subsubsection*{Scalar DM, scalar mediator:}
\begin{align}
\sigma_{{\rm DM}\text{--}\nu}&\simeq \frac{g^2g'^2E_\nu^2}{8\pi m_{\rm DM}^2m_{\rm \phi}^4}, \\
g&\lesssim 2\times 10^{-4}\ \left(\frac{g'}{{\rm MeV}} \right)^{-1} \left(\frac{m_{\rm DM}}{\rm MeV} \right)^{3/2}\left(\frac{m_\phi}{\rm MeV} \right)^2\left(\frac{E_\nu}{E_\nu^0} \right)^{-1}\left(\frac{\sigma_{{\rm DM}\text{--}\nu}/m_{\rm DM}}{10^{-49}\ {\rm cm^2/ {\rm MeV}}} \right)^{1/2}.
\label{SDM_Smed}
\end{align}

\subsubsection*{Dirac fermion DM, scalar/vector mediator:}
\begin{align}
\sigma_{{\rm DM}\text{--}\nu}&\simeq \frac{g^2g'^2E_\nu^2}{2\pi m_{\rm \phi}^4}, \\
g&\lesssim 8\times 10^{-5}\ \left(\frac{g'}{1} \right)^{-1} \left(\frac{m_{\rm DM}}{\rm MeV} \right)^{1/2}\left(\frac{m_\phi}{\rm MeV} \right)^2\left(\frac{E_\nu}{E_\nu^0} \right)^{-1}\left(\frac{\sigma_{{\rm DM}\text{--}\nu}/m_{\rm DM}}{10^{-49}\ {\rm cm^2/ {\rm MeV}}} \right)^{1/2}.
\label{DDM_Smed}
\end{align}

\subsubsection*{Scalar DM, fermion mediator:}
\begin{align}
\sigma_{{\rm DM}\text{--}\nu}&\simeq \frac{2g^4E_\nu^4m_{\rm DM}^2}{\pi m_{\rm \phi}^8}, \\
g&\lesssim 0.3\ \left(\frac{m_{\rm DM}}{\rm MeV} \right)^{-1/4}\left(\frac{m_\phi}{\rm MeV} \right)^2\left(\frac{E_\nu}{E_\nu^0} \right)^{-1} \left(\frac{\sigma_{{\rm DM}\text{--}\nu}/m_{\rm DM}}{7 \times 10^{-62}\ {\rm cm^2/ {\rm MeV}}} \right)^{1/4}.
\label{SDM_fmed}
\end{align}

\subsubsection*{Scalar DM, fermion mediator with $m_{\rm DM}=m_\phi$:}
\begin{align}
\sigma_{{\rm DM}\text{--}\nu}&\simeq \frac{g^4}{8\pi m_{\rm DM}^2}, \\
g&\lesssim 4\times 10^{-4}\ \left(\frac{m_{\rm DM}}{\rm MeV} \right)^{3/4}\left(\frac{\sigma_{{\rm DM}\text{--}\nu}/m_{\rm DM}}{4 \times 10^{-37}\ {\rm cm^2/ {\rm MeV}}} \right)^{1/4}.
\label{SDM_fmed_eq}
\end{align}

\section{Discussions}
\label{sec6}


\subsection{Uncertainties of constraints}
\label{sec6.1}

We have used the observational data of 270 Milky-Way satellites after completeness corrections \cite{DES:2019vzn}.
A companion paper \cite{DES:2019ltu} in ref.~\cite{DES:2019vzn} predicts $220\pm 50$ Milky-Way satellites after completeness corrections at $68\%$ confidence level. 
Adopting 170 Milky-Way satellites, we find that our constraints on the DM-neutrino cross section with a satellite forming condition of $m_a>10^8M_\odot$ become weaker by a factor of $3, 40, 6\times 10^2$ for $n=0,2,4$, respectively. 
Theoretical uncertainties of the completeness corrections are marginalized except for the Milky-Way halo mass. The uncertainty of the Milky-Way halo mass is expected to change the number of the total satellites from 170 to 280 \cite{DES:2019ltu}, based on the Gaia data of the Milky-Way halo mass.

Our subhalo model coincides with N-body simulations for the ETHOS models with DM-DR interactions within a factor of 1.8 as in figure~\ref{fig:ETHOS_subhalo}, but the result of cumulative number of satellites on small scales of $V_{\rm max}\lesssim 10\ {\rm km/s}$ by our model is in better agreement with the N-body simulations than a factor of 1.8. 
This factor might also induce the same order of uncertainties as the above uncertainties.

\subsection{Comparison with results from astrophysical neutrinos}
\label{sec6.2}

Observations of astrophysical neutrinos propagating in DM also constrain the scattering cross sections. Due to the different energy scales of neutrinos, there is no simple mapping to the cross section between cosmological constraints and astrophysical constraints of the DM-neutrino cross sections.

In the following, we consider mainly the DM scattering with high-energy neutrinos \cite{Arguelles:2017atb,Pandey:2018wvh,Alvey:2019jzx,Choi:2019ixb,Murase:2019xqi,Cline:2022qld,Ferrer:2022kei,Cline:2023tkp}.
We will comment on the constraints from observations of SN 1987A neutrinos in the final paragraph of this section.
The most stringent constraints from high energy neutrinos are imposed by neutrino events with $E_\nu\sim 10\ {\rm TeV}$ from active galaxy NGC 1068.
The upper bounds of the scattering cross section from these events are roughly $\sigma_{\rm DM\text{--}\nu,n}\lesssim 10^{-30}\ {\rm cm^2}\ (m_{\rm DM}/{\rm GeV})(E_\nu/10\ {\rm TeV})^n$ \cite{Cline:2023tkp}.\footnote{The authors assume DM particles do not annihilate today as in the scenario of asymmetric DM. If DM annihilates, the density of halos through which neutrinos propagate is reduced, weakening this constraint.}

We consider several types of scattering between DM and neutrinos via a mediator particle, following appendix~C of ref.~\cite{Arguelles:2017atb}, and compare with our constraints derived in section~\ref{sec5.2}. For scattering of DM with high-energy neutrinos, we consider the parameters of $E_\nu \gg m_\phi \gtrsim m_{\rm DM}$, where $m_\phi$ and $m_{\rm DM}$ are the mediator and DM masses, respectively. For $m_\phi\gg E_\nu$ and/or $m_{\rm DM}\gg E_\nu$, the cross section would be further suppressed by their masses, which is of less interest. We also leave the case of $m_\phi \gg m_{\rm DM}$ as future work for simplicity. See ref.~\cite{Arguelles:2017atb} for a discussion of comparison in the general case.
As a result, the constraints on the DM-neutrino scattering cross section from cosmology and high-energy neutrinos are highly complementary.

\subsubsection*{Scalar DM, scalar mediator:}
\begin{align}
\sigma_{{\rm DM}\text{--}\nu}&\simeq \frac{g^2g'^2}{64\pi E_\nu^2m_{\rm DM}^2}\log\left(\frac{2E_\nu m_{\rm DM}}{m_\phi^2} \right),  \\
g&\lesssim 5\times 10^1\ \left(\frac{g'}{\rm MeV}\right)^{-1}\left(\frac{m_{\rm DM}}{\rm MeV} \right)^{3/2}\left(\frac{E_\nu}{10\ {\rm TeV}} \right)\left(\frac{\sigma_{{\rm DM}\text{--}\nu}/m_{\rm DM}}{10^{-33}\ {\rm cm^2}/{\rm MeV}} \right)^{1/2},
\end{align}
where we assume the log term is $\sim \mathcal{O}(10)$.
The constraint from high energy neutrinos is weaker than that from the Milky-Way satellite observations, eq.~(\ref{SDM_Smed}).

\subsubsection*{Dirac fermion DM, scalar mediator:}
\begin{align}
\sigma_{{\rm DM}\text{--}\nu}&\simeq \frac{g^2g'^2}{32\pi E_\nu m_{\rm DM}},  \\
g&\lesssim 5\times 10^{-2}\ \left(\frac{g'}{1} \right)^{-1}\left(\frac{m_{\rm DM}}{\rm MeV} \right)\left(\frac{E_\nu}{10\ {\rm TeV}} \right)^{1/2}\left(\frac{\sigma_{{\rm DM}\text{--}\nu}/m_{\rm DM}}{10^{-33}\ {\rm cm^2/MeV}} \right)^{1/2}.
\end{align}
The above constraint is weaker than that from the Milky-Way satellite observations, eq.~(\ref{DDM_Smed}).

\subsubsection*{Dirac fermion DM, vector mediator:}
\begin{align}
\sigma_{{\rm DM}\text{--}\nu}&\simeq \frac{g^2g'^2}{4\pi m_{\rm \phi}^2},  \\
g&\lesssim 6\times 10^{-6}\ \left(\frac{g'}{1} \right)^{-1}\left(\frac{m_{\phi}}{{\rm MeV}} \right)\left(\frac{m_{\rm DM}}{{\rm MeV}} \right)^{1/2}\left(\frac{\sigma_{{\rm DM}\text{--}\nu}/m_{\rm DM}}{10^{-33}\ {\rm cm^2/MeV}} \right)^{1/2}.
\label{HE_DDM_Smed}
\end{align}
The above constraint is stronger than that from the Milky-Way satellite observations, eq.~(\ref{DDM_Smed}) for $m_\phi\gtrsim m_{\rm DM}\gtrsim 1\ {\rm MeV}$.
For $100\ {\rm keV}\gtrsim m_\phi \gtrsim m_{\rm DM}$, eq.~(\ref{HE_DDM_Smed}) is comparable or weaker than eq.~(\ref{DDM_Smed}).

\subsubsection*{Scalar DM, fermion mediator:}
\begin{align}
\sigma_{{\rm DM}\text{--}\nu}&\simeq \frac{g^4}{32\pi E_\nu m_{\rm DM}}\log\left(\frac{2E_\nu m_{\rm DM}}{m_\phi^2} \right), \\
g&\lesssim 0.1\ \left(\frac{m_{\rm DM}}{{\rm MeV}} \right)^{1/2}\left(\frac{E_\nu}{10\ {\rm TeV}} \right)^{1/4} \left(\frac{\sigma_{{\rm DM}\text{--}\nu}/m_{\rm DM}}{10^{-33}\ {\rm cm^2/MeV}} \right)^{1/4},
\label{HE_SDM_fmed}
\end{align}
where we assume the log term is $\sim \mathcal{O}(10)$.
The above constraint is stronger than that from the Milky-Way satellite observations, eq.~(\ref{SDM_fmed}), for $m_\phi \gtrsim m_{\rm DM}\gtrsim 1\ {\rm MeV}$.
For $100\ {\rm keV}\gtrsim m_\phi \gtrsim m_{\rm DM}$, eq.~(\ref{HE_SDM_fmed}) is weaker than eq.~(\ref{SDM_fmed}).
In addition, if $m_{\phi}=m_{\rm DM}$, eq.~(\ref{HE_SDM_fmed}) is much weaker than that from the Milky-Way satellite observations, eq.~(\ref{SDM_fmed_eq}).

The constraints are also provided by observations of neutrinos from SN 1987A. The thickness of DM through which SN 1987A neutrinos propagate is $\int_{\rm 8\ kpc}^{\rm 50\ kpc} \rho(r) dr\sim 10^{25}\ {\rm MeV\ cm^{-2}}$, where $\rho(r)$ is the DM density in the Milky Way.
Then the constraints from SN 1987A observations read $\sigma_{\rm DM\text{--}\nu,n}\lesssim 10^{-25}\ {\rm cm^2}(m_{\rm DM}/{\rm MeV})(E_\nu/10\ {\rm MeV})^n$ \cite{Mangano:2006mp}, where
the average energy of SN 1987A neutrinos is $\sim 10\ {\rm MeV}$. In many of the parameter regions of the above examples, the constraints on the couplings from SN 1987A neutrinos are comparable or weaker than those from the Milky-Way satellite or high-energy neutrino observations.


\subsection{Comparison with results from the Milky-Way satellites in the literature}
\label{sec6.3}

 Strictly speaking, there is no previous constraint on the DM-neutrino scattering from the Milky-Way satellite observations, but refs.~\cite{Boehm:2014vja,Escudero:2018thh} provided the constraints on the DM-photon scattering cross section, which might be similar to the limits on the DM-neutrino cross section.
 We compare our methods and results with refs.~\cite{Boehm:2014vja,Escudero:2018thh}.

 In ref.~\cite{Boehm:2014vja}, using the N-body simulations of subhalos and the kinematics data of 40 Milky-Way satellites with $V_{\rm max}\geq 8 {\rm km/s}$, the authors constrain the DM-photon constant cross section of $\sigma_{{\rm DM}\text{--}\gamma}<2\times 10^{-33}\ {\rm cm^2}\ (m_{\rm DM}/{\rm GeV})$ for the Milky-Way mass of $2\times 10^{12}M_\odot$.
 In the numerical simulations of ref.~\cite{Boehm:2014vja}, the maximum circular velocity, $V_{\rm max}$, is calculated by the stellar line-of-sight velocity dispersion, $V_{\rm max}=\sqrt{3}\sigma$, which may be underestimated.
Then our model predicts a larger number of subhalos than the N-body simulations of ref.~\cite{Boehm:2014vja}, weakening the constraints on the DM-radiation scattering cross section.

In ref.~\cite{Escudero:2018thh}, using the data of ref.~\cite{Newton:2017xqg} and a semi-analytical model with a satellite galaxy formation condition, the authors constrain the DM-photon constant cross section of $\sigma_{{\rm DM}\text{--}\gamma}<3.3\times 10^{-34}\ {\rm cm^2}$ for the Milky-Way mass of $2\times10^{12}M_\odot$.
The constraints from our model are weaker than those of ref.~\cite{Escudero:2018thh}, predicting a larger number of subhalos. In addition, our model can calculate the number of subhalos as a function of various quantities while the method of ref.~\cite{Escudero:2018thh} can only calculate that as a function of subhalo mass. Our semi-analytical model is more flexible.

\section{Conclusions}
\label{sec7}

In this work, we study a semi-analytical subhalo model for interacting dark matter scenarios, which induce dark acoustic oscillations, based on the EPS formalism including subhalos' tidal evolution. Using this semi-analytical model, we have provided stringent upper bounds on the elastic scattering of DM and neutrinos from the latest population data of the Milky-Way satellite galaxies by DES and PS1 surveys.

The modifications to our semi-analytical model compared with the model for WDM \cite{Dekker:2021scf} are only using the smooth-k filter and changing the critical overdensity.
Our model is in a reasonable agreement with the publicly available results of N-body simulations for three DM-DR interaction scenarios. 
This agreement indicates that our model is a simple and powerful tool to study subhalo properties in presence of DAO induced by DM-radiation interactions quickly and comprehensively.

By comparing the data of the Milky-Way satellites from DES and PS1 surveys with the results from our model,
we have obtained the constraints on the DM-neutrino scattering cross section of $\sigma_{{\rm DM}\text{--}\nu,n}\propto E_\nu^n$ $(n=0,2,4)$ at $95\%$ CL, $\sigma_{{\rm DM}\text{--}\nu,0}< 10^{-32}\ {\rm cm^2}\ (m_{\rm DM}/{\rm GeV})$, $\sigma_{{\rm DM}\text{--}\nu,2}< 10^{-43}\ {\rm cm^2}\ (m_{\rm DM}/{\rm GeV})(E_\nu/E_{\nu}^0)^2$ and $\sigma_{{\rm DM}\text{--}\nu,4}< 10^{-54}\ {\rm cm^2}\ (m_{\rm DM}/{\rm GeV})(E_\nu/E_{\nu}^0)^4$.
By imposing a satellite forming condition, we obtain the upper bounds on the DM-neutrino cross section at $95\%$ CL, $\sigma_{{\rm DM}\text{--}\nu,0}< 4\times 10^{-34}\ {\rm cm^2}\ (m_{\rm DM}/{\rm GeV})$, $\sigma_{{\rm DM}\text{--}\nu,2}< 10^{-46}\ {\rm cm^2}\ (m_{\rm DM}/{\rm GeV})(E_\nu/E_{\nu}^0)^2$ and $\sigma_{{\rm DM}\text{--}\nu,4}< 7\times 10^{-59}\ {\rm cm^2}\ (m_{\rm DM}/{\rm GeV})(E_\nu/E_{\nu}^0)^4$.
 The results obtained in this analysis are independent of specific particle physics models.
 In section~\ref{sec5.2}, we have mapped their constraints to couplings for specific DM-neutrino interactions.


\section*{Acknowledgments}
KA is supported by IBS under the project code, IBS-R018-D1.
The work of SA was supported by MEXT KAKENHI Grant Numbers, JP20H05850 and JP20H05861.

\appendix

\section{Fitting formula of critical overdensity}
\label{appa}

DM-radiation interactions may affect a critical overdensity for DM collapse, $\delta_c$, due to heating DM and producing pressure for DM particles that prevents their collapse.
The collapse of DM with pressure is characterized by the Jeans mass, which is the halo mass for which pressure balances gravity in the linear regime.
In the following, we consider DM-neutrino interactions. It is straightforward to apply the following discussions to other DM-radiation interaction scenarios.

DM decouples with neutrinos when $\mathcal{H}\sim \Gamma_{{\rm DM\text{--}}\nu}$. The decoupling redshift $z_{\rm dec}$ is given by
\begin{align}
1+z_{\rm dec}=\left[4.5\times 10^{12}\left(\frac{u_{{\rm DM}\text{--}\nu,n}^0}{10^{-7}} \right)^{-1} \right]^{\frac{1}{2+n}}.
\label{zdec}
\end{align}
The temperature of DM is obtained by, assuming DM is non-relativistic at the decoupling,
\begin{align}
T_{\rm DM}(z)&=T_{\nu}^0\frac{(1+z)^2}{1+z_{\rm dec}},
\end{align}
where $T_{\nu}^0=1.945\ {\rm K}$ is the temperature of relic cosmic neutrinos today. The Jeans mass in the interacting DM scenario is defined as
\begin{align}
M_{J}^{{\rm IDM}}&\equiv \frac{4\pi}{3}\bar{\rho}\left(\frac{\lambda_J^{{\rm IDM}}}{2} \right)^3, \\
\lambda_J^{{\rm IDM}}&=\left(\frac{\pi \gamma T_{\rm DM}}{G \bar{\rho} m_{\rm DM}} \right)^{1/2},
\end{align}
where $\gamma=5/3$ for DM particles and $G$ is the gravitational constant. 
The Jeans mass can be written,
\begin{align}
M_{J}^{{\rm IDM}}=1.3\times 10^6 M_\odot \left(\frac{1+z}{3000} \right)^{3/2}\left(\frac{T_\nu^0}{2\ {\rm K}} \right)^{3/2}\left(\frac{{\rm MeV}}{m_{\rm DM}} \right)^{3/2}\left(\frac{1+z_{\rm dec}}{10^6}\right)^{-3/2}.
\end{align}
After the matter-radiation equality of $z_{\rm eq}\sim 3000$, the density fluctuations begin to grow as $\propto a$. 

In ref.~\cite{Barkana:2001gr}, the authors discussed the overcritical density for WDM. Using a model for WDM, where WDM is an adiabatic gas with a gas temperature, $T\propto a^{-2}$, they solved one-dimensional hydrodynamical simulations for spherical collapse with pressure (i.e., with the gas temperature $T$). 
The hydrodynamical equations for the WDM model are mathematically identical to those for the DM-radiation interaction scenario. Thus, we can apply the results in ref.~\cite{Barkana:2001gr} to our model. The fitting formula of the critical overdensity for the DM-radiation scenario is, based on refs.~\cite{Barkana:2001gr} and \cite{Benson:2012su},
\begin{align}
\delta_{c,{{\rm IDM}}}(M,z)=\delta_{c, {\rm CDM}}(t)\left\{ h(x)\frac{0.04}{\exp(2.3x)}+[1-h(x)]\exp \left[\frac{0.31687}{\exp(0.809x)} \right] \right\},
\label{deltac_fit}
\end{align}
where $x=\log[M/M_J^{{\rm IDM}}(z_{\rm eq})]$ and
$h(x)=(1+\exp[(x+2.4)/0.1])^{-1}$.
For $x\gg 1$, we obtain $\delta_{c,{{\rm IDM}}}(M,z)\simeq \delta_{c, {\rm CDM}}(t)$.
In section \ref{sec5}, we consider the number of Milky-Way satellites $N_{\rm sh}$ with $V_{\rm max}\geq 4\ {\rm km/s}$ or $m_a\geq 10^8M_\odot$.
The relation between the subhalo mass $m$ and their maximal circular velocity $V_{\rm max}$ in presence of the DM-radiation interactions is still non-trivial, but in the WDM case, which has a truncated power spectrum, $V_{\rm max}\sim 8\ {\rm km/s}$ corresponds to $m\sim 10^8M_\odot$ \cite{Lovell:2013ola}.
In the interaction strength of our interest with $m_{\rm DM}\gtrsim {\rm MeV}$, we typically obtain $M_J(z_{\rm eq})\lesssim 10^6 M_\odot$. To compute $N_{\rm sh}$ with $V_{\rm max}\geq 4\ {\rm km/s}$ or $m_a\geq 10^8M_\odot$, we expect to be able to approximate
\begin{align}
\delta_{c,{{\rm IDM}}}(M,z)\simeq \delta_{c, {\rm CDM}}(t).
\label{delc_CDM}
\end{align}
We have confirmed  this approximation does not affect $N_{\rm sh}$ with $V_{\rm max}\geq 4\ {\rm km/s}$ or $m_a\geq 10^8M_\odot$ for $m_{\rm DM} \geq {\rm MeV}$ and $u_{{\rm DM}\text{--}\nu,0}^0=10^{-6}$. To compute the properties of smaller subhalos or lighter DM particles, we should use the fitting formula of the critical overdensity given by eq.~(\ref{deltac_fit}). Eq.~(\ref{delc_CDM}) is a conservative choice, where the number of subhalos is less suppressed, compared with eq.~(\ref{deltac_fit}).


\section{Comparison with other parametrizations of the smooth-k filter}
\label{appb}

In our fiducial model, we adopt the smooth-k filter with $(c,\beta)=(3.7,3.5)$ \cite{Sameie:2018juk,Bohr:2021bdm}.
We will compare with other choices of $(c,\beta)=(3.3,3)$ \cite{Schaeffer:2021qwm} and $(c,\beta)=(3.3,4.8)$ \cite{Leo:2018odn}, which are calibrated by simulations of the halo abundance in the universe for other DAO scenarios.

\begin{figure}
 \begin{minipage}{0.5\hsize}
  \begin{center}
   \includegraphics[width=80mm]{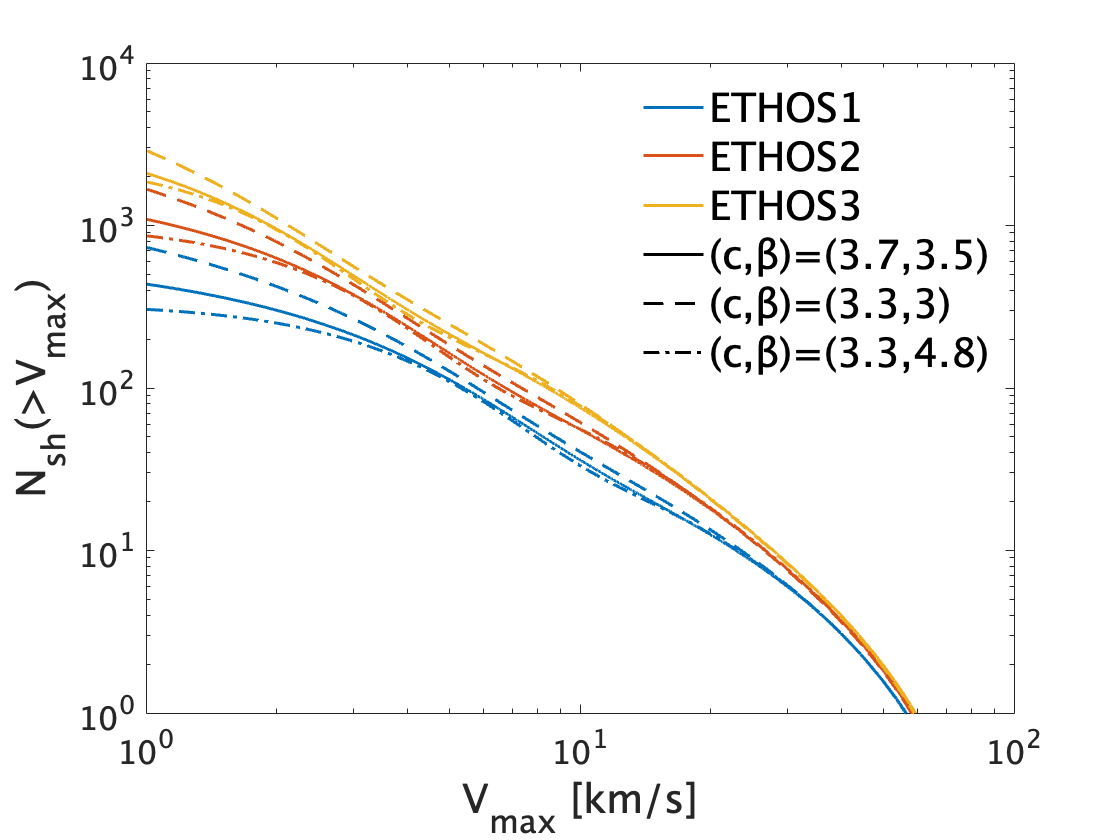}
  \end{center}
 \end{minipage}
 \begin{minipage}{0.5\hsize}
 \begin{center}
  \includegraphics[width=80mm]{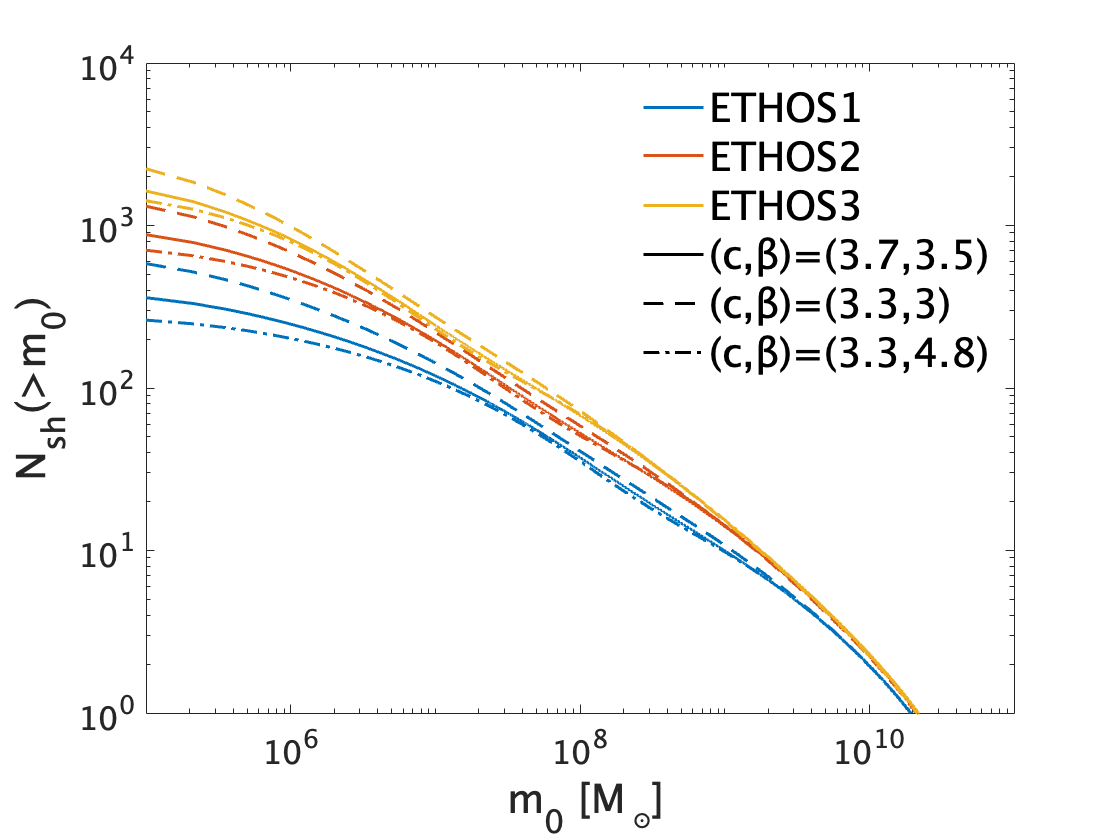}
 \end{center}
 \end{minipage} 
 \caption{Cumulative number of subhalos as a function of the maximum circular velocity (left) and the virial subhalo mass (right) with the Milky-Way mass of $M_{200}=1.6\times 10^{12}M_\odot$ for ETHOS models described in ref.~\cite{Vogelsberger:2015gpr}.
 Solid, dashed and dot-dashed lines correspond to our models with the smooth-k filter of parameter values $(c,\beta)=(3.7,3.5)$\cite{Sameie:2018juk,Bohr:2021bdm}, $(3.3, 3)$\cite{Schaeffer:2021qwm} and $(3.3, 4.8)$\cite{Leo:2018odn}, respectively.}
    \label{fig:ETHOS_cbeta}
\end{figure}

In figure~\ref{fig:ETHOS_cbeta}, we show the number of subhalos predicted by our models as a function of the maximum circular velocity $V_{\rm max}$ (left panels) and the virial subhalo mass $m_0$ (right panels) for ETHOS models. Solid, dashed and dot-dashed lines denote subhalo abundances prediced by the models with $(c,\beta)=(3.7,3.5),\ (3.3,3)$ and $(3.3,4.8)$, respectively.
All models predict almost the same results.
Since the different choices of $(c,\beta)$ are calibrated by simulations of the halo abundance for different DAO scenarios, these results indicate that our fiducial model with $(c,\beta)=(3.7,3.5)$ can be applied to a broad class of DM-radiation interaction scenarios.

\section{Distributions of the number of subhalos}
\label{appc}
\begin{figure}
 \begin{minipage}{0.5\hsize}
  \begin{center}
   \includegraphics[width=80mm]{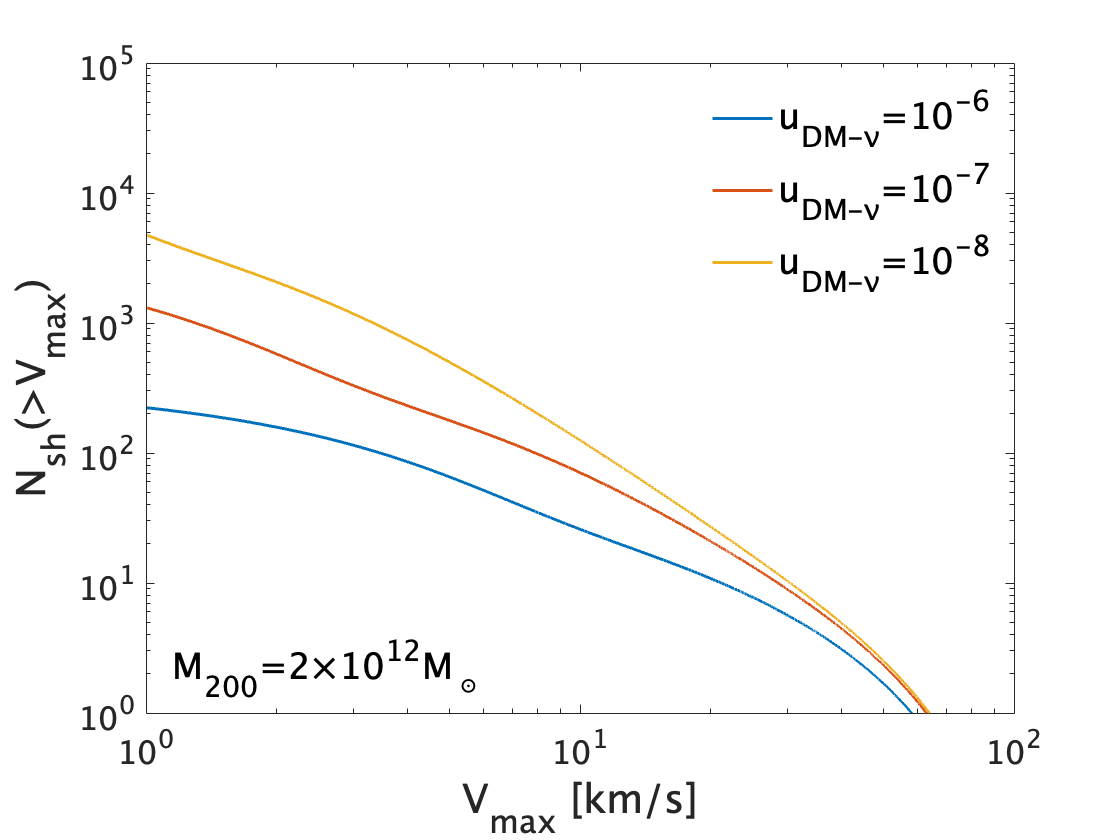}
  \end{center}
 \end{minipage}
 \begin{minipage}{0.5\hsize}
 \begin{center}
  \includegraphics[width=80mm]{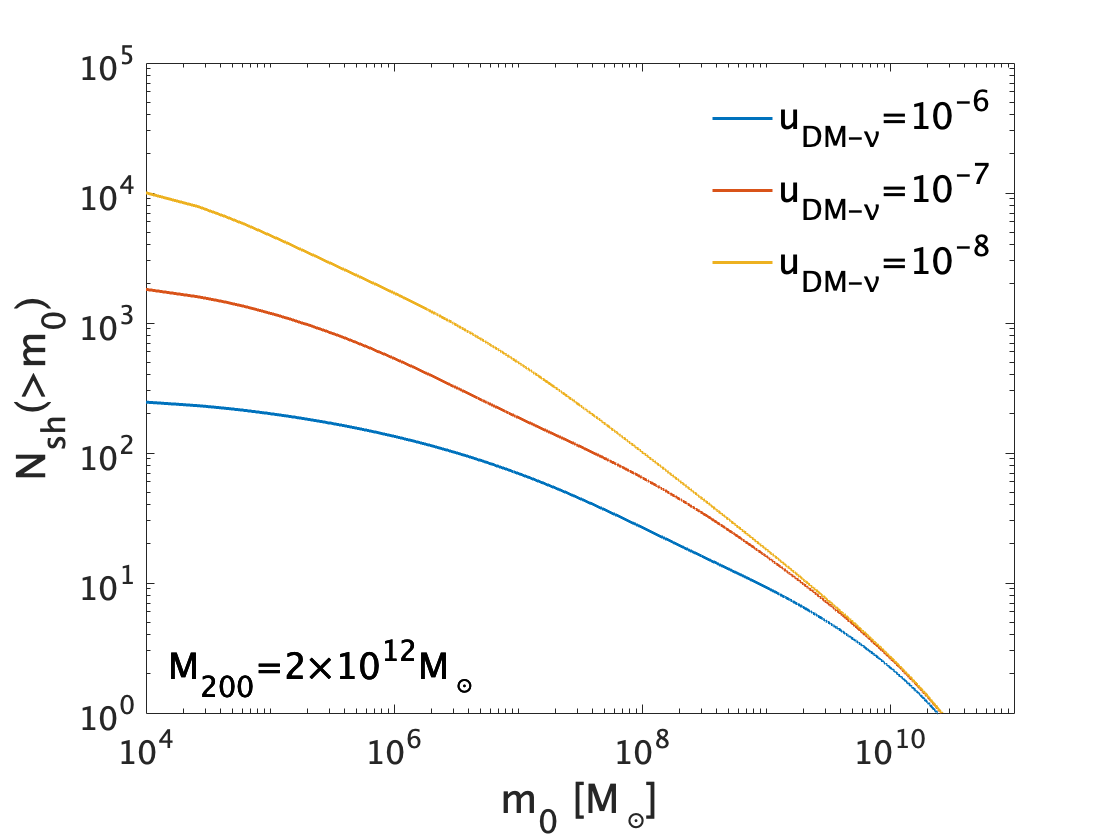}
 \end{center}
 \end{minipage} \\
 \begin{minipage}{0.5\hsize}
 \begin{center}
  \includegraphics[width=80mm]{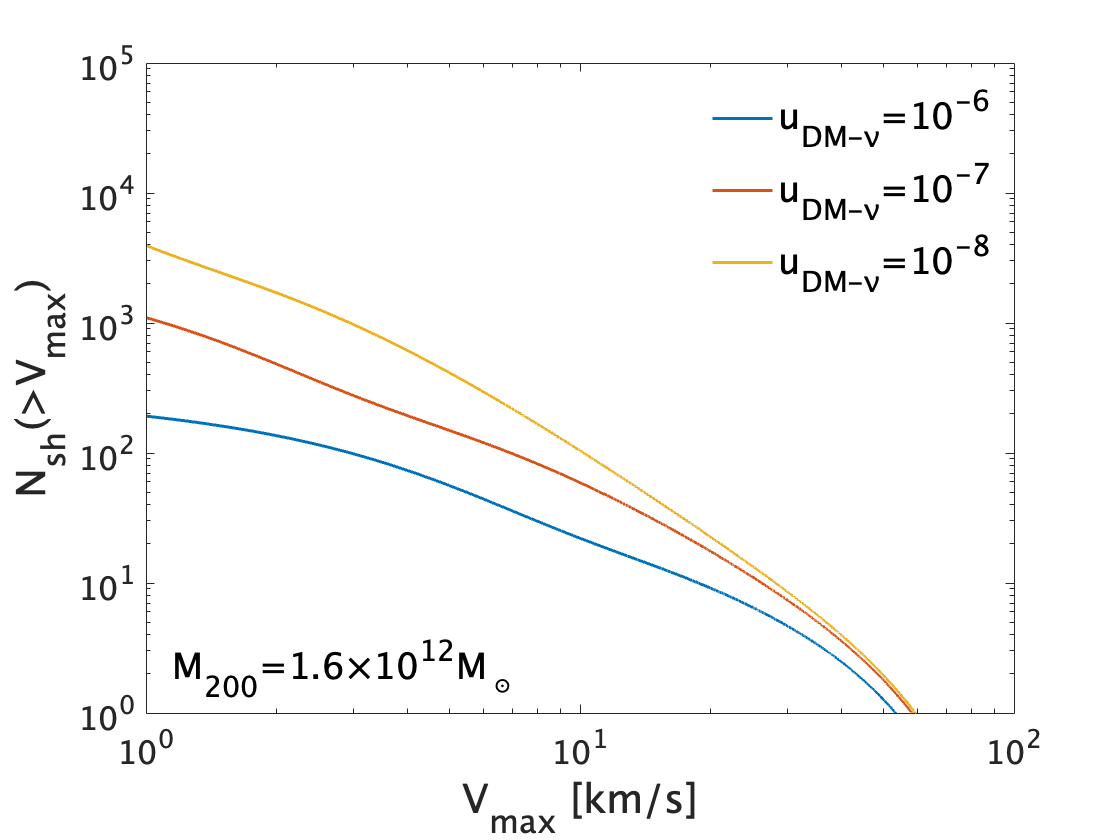}
 \end{center}
 \end{minipage}
 \begin{minipage}{0.5\hsize}
 \begin{center}
  \includegraphics[width=80mm]{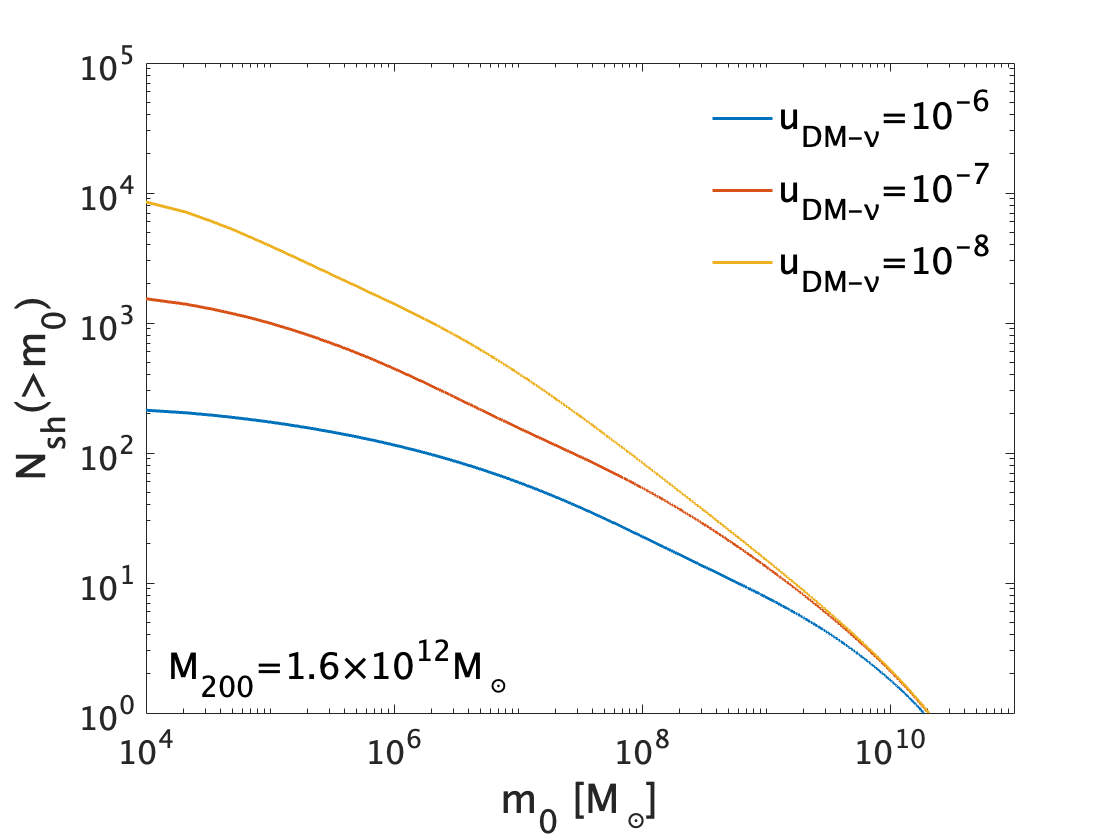}
 \end{center}
 \end{minipage} \\
 \begin{minipage}{0.5\hsize}
 \begin{center}
  \includegraphics[width=80mm]{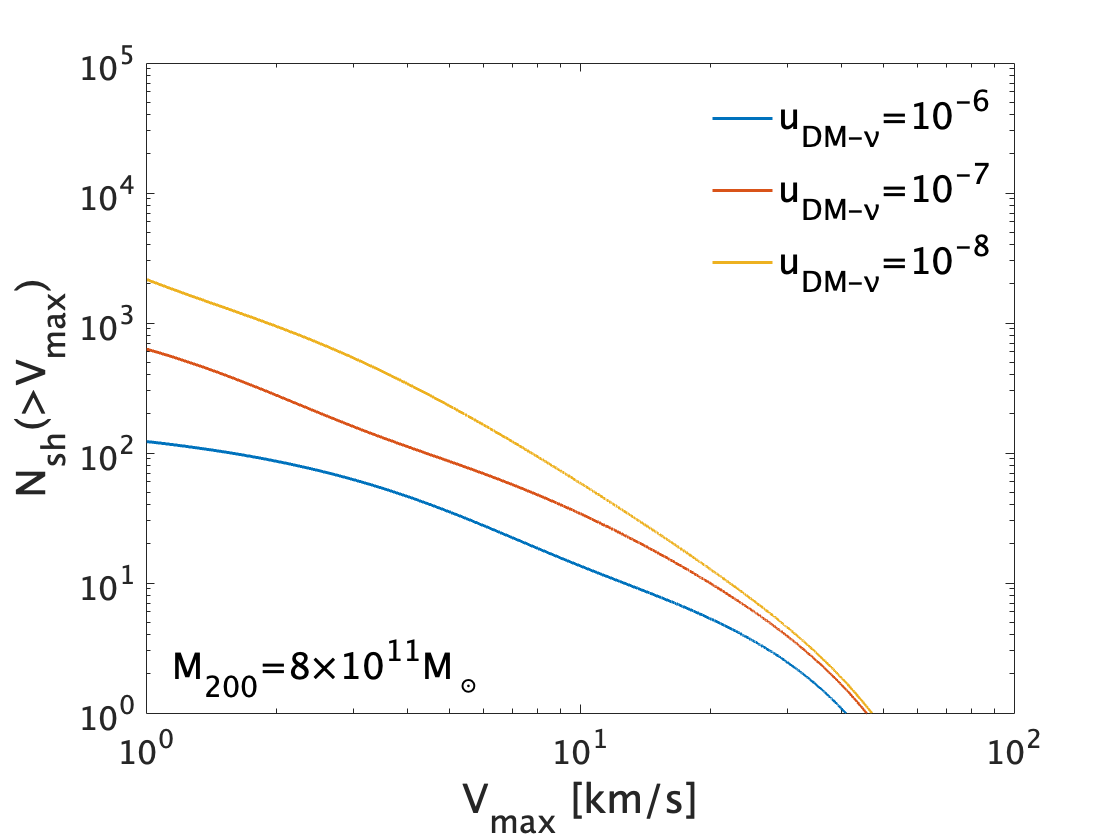}
 \end{center}
 \end{minipage}
 \begin{minipage}{0.5\hsize}
 \begin{center}
  \includegraphics[width=80mm]{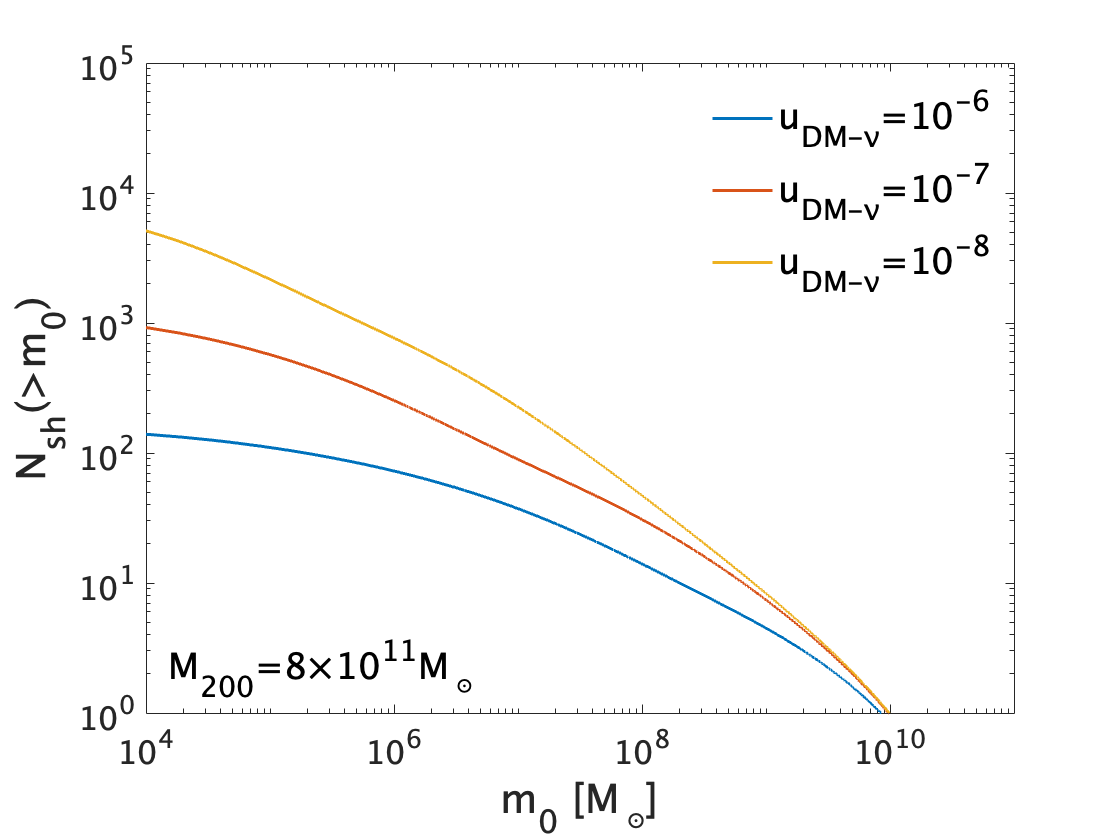}
 \end{center}
 \end{minipage}
 \caption{Cumulative number of subhalos as a function of the maximum circular velocity (left) and the virial subhalo mass (right) with the Milky-Way mass of $M_{200}=2\times 10^{12}M_\odot$ (top), $1.6\times 10^{12}M_\odot$ (middle) and $8\times 10^{11}M_\odot$ (bottom). We consider three DM-neutrino interaction scenarios with the constant DM-neutrino scattering cross section of $u_{{\rm DM}\text{--}\nu}=10^{-6}$ (blue lines), $u_{{\rm DM}\text{--}\nu}=10^{-7}$ (red lines) and $u_{{\rm DM}\text{--}\nu}=10^{-8}$ (yellow lines).}
    \label{fig:Dist_DM_Nu}
\end{figure}

In figure~\ref{fig:Dist_DM_Nu}, we show the number of subhalos predicted by our model as a function of the maximum circular velocity $V_{\rm max}$ (left panels) and the virial subhalo mass $m_0$ (right panels) for several examples. In this figure, we consider the constant DM-neutrino cross section for simplicity. We take $u_{{\rm DM}\text{--}\nu}=10^{-6}$ (blue lines), $u_{{\rm DM}\text{--}\nu}=10^{-7}$ (red lines) and $u_{{\rm DM}\text{--}\nu}=10^{-8}$ (yellow lines). In each panel, we assume the Milky-Way mass of $M_{200}=2.0\times 10^{12}M_\odot$ (top panels), $M_{200}=1.6\times 10^{12}M_\odot$ (middle panels) and $M_{200}=0.8\times 10^{12}M_\odot$ (bottom panels).

\bibliographystyle{JHEP}
\bibliography{reference}

\end{document}